\documentclass[aps,pre,preprint, showpacs, superscriptaddress]{revtex4}

\usepackage[dvips]{graphicx}
\usepackage{latexsym}
\usepackage{natbib}

\begin{document}

\bibliographystyle{apsrev}

\title{Optimal Paths in Complex Networks with
  Correlated Weights:\\
  The World-wide Airport Network}

\author{Zhenhua Wu}
\affiliation{Center for Polymer Studies, Boston University, Boston,
  Massachusetts 02215, USA}
\author{Lidia A. Braunstein}
\affiliation{Center for Polymer Studies, Boston University, Boston,
  Massachusetts 02215, USA}
\affiliation{Departamento de F\'{\i}sica, Facultad de Ciencias Exactas y
  Naturales, Universidad Nacional de Mar del Plata, Funes 3350, 7600 Mar del
  Plata, Argentina}
\author{Vittoria Colizza}
\affiliation{School of Informatics and Department of Physics, Indiana University,
  Bloomington, Indiana 47406, USA}
\author{Reuven Cohen}
\author{Shlomo Havlin}
\affiliation{Minerva Center and Department of Physics, Bar-Ilan
  University, Ramat Gan, Israel}
\author{H. Eugene Stanley}
\affiliation{Center for Polymer Studies, Boston University, Boston,
  Massachusetts 02215, USA}

\today

\begin{abstract}

We study complex networks with weights, $w_{ij}$, associated with each link
connecting node $i$ and $j$. The weights are chosen to be correlated with the
network topology in the form found in two real world examples, (a) the
world-wide airport network, and (b) the {\it E. Coli} metabolic network.
Here $w_{ij} \sim x_{ij} (k_i k_j)^\alpha$, where $k_i$ and $k_j$ are the
degrees of nodes $i$ and $j$, $x_{ij}$ is a random number and $\alpha$
represents the strength of the correlations. The case $\alpha > 0$ represents
correlation between weights and degree, while $\alpha < 0$ represents
anti-correlation and the case $\alpha = 0$ reduces to the case of no
correlations. We study the scaling of the lengths of the optimal paths,
$\ell_{\rm opt}$, with the system size $N$ in strong disorder for scale-free
networks for different $\alpha$. We find two different universality classes
for $\ell_{\rm opt}$ in strong disorder depending on $\alpha$: (i) if $\alpha
> 0$, then for $\lambda > 2$ the scaling law $\ell_{\rm opt} \sim N^{1/3}$,
where $\lambda$ is the power-law exponent of the degree distribution of
scale-free networks, (ii) if $\alpha \le 0$, then $\ell_{\rm opt} \sim
N^{\nu_{\rm opt}}$ with $\nu_{\rm opt}$ identical to its value for the
uncorrelated case $\alpha = 0$. We calculate the robustness of correlated
scale-free networks with different $\alpha$, and find the networks with
$\alpha < 0$ to be the most robust networks when compared to the other values
of $\alpha$. We propose an analytical method to study percolation phenomena
on networks with this kind of correlation, and our numerical results suggest
that for scale-free networks with $\alpha < 0$, the percolation threshold,
$p_c$ is finite for $\lambda > 3$, which belongs to the same universality
class as $\alpha = 0$. We compare our simulation results with the real
world-wide airport network, and we find good agreement.

\end{abstract}



\maketitle

\section{Introduction}

Recently attention has focused on the topic of complex networks, which
characterize many natural and man-made systems, such as the internet, airline
transport system, power grid infrastructures, biological and social
interaction systems~\cite{Barabasi_rmp_review, vespignani_book,
Mendes_book}. Network structure systems are visualized by nodes representing
individuals, organizations, or computers and by links between them
representing their interactions. Significant topological features were
discovered, such as the clustering and small-world
properties~\cite{Watts_smallworld}. There exists evidence that many real
networks possess a scale-free (SF) degree distribution characterized by a
power law tail given by ${\cal P}(k) \sim k^{-\lambda}$, where $k$ is the
degree of a node and $\lambda$ measures the broadness of the
distribution~\cite{Barabasi_sf}. In most studies, all links or nodes in the
network are regarded as identical. Thus the topological structure of the
network determines all the other properties of the network, such as
robustness, percolation threshold~\cite{Cohen_robust, Callaway}, the average
shortest path length, $\ell_{\rm min}$~\cite{Cohen_ultrasmall} and
transport~\cite{lopez_prl}.

In many real world networks the links are not equally weighted. For example,
the links between computers in the Internet network have different capacities
or bandwidths and the airline network links between different pairs of cities
have different numbers of passengers. To better understand real networks,
several studies have been carried out on weighted network
models~\cite{Park_betweenness, Barrat_prl, Yook_weightEvolve, Dafang_wn,
Wenxu_prl, Cieplak:op1, Porto, Lidia_op_prl, Strelniker, zhenhua}. For
weighted networks, an important quantity that characterizes the information
flow is the optimal path, which is the path that minimizes the total weight
along the path~\cite{Cieplak:op1, Porto, Lidia_op_prl, Strelniker, zhenhua,
Sameet_mst, tomer}. Optimal paths play important roles in many dynamic
systems such as current flow of random resistor networks, where the dynamics
of current flow is strongly controlled by the optimal path~\cite{Strelniker,
zhenhua}. The above studies assume that the weights on the links are purely
random --- i.e. uncorrelated with the topology of the network.

However, recently several studies on networks with weights on the links, such
as the world-wide airport networks (WAN) and the {\it E. coli} metabolic
networks~\cite{Barrat_pnas, Almaas_ecoli, Macdonald_ecoli} show that the
weights are correlated with the network topology~\cite{Barrat_pnas,
Macdonald_ecoli}. For the WAN, each link is a direct flight between two
airports $i$ and $j$ and the weight $w_{ij}$ is the number of passengers
between them during a period of time. The mean traffic on links can be
characterized by: $\langle T_{ij} \rangle \sim \langle k_i k_j
\rangle^{\theta}$, where $k_i$ and $k_j$ are the degrees of nodes $i$ and $j$
and $\theta > 0$~\cite{Macdonald_ecoli}.

In this paper, we study how the correlations between the topology and the
weights affect the robustness, the percolation threshold, the scaling of the
optimal path length and the minimum spanning tree by studying the weighted SF
model. The WAN was found to be a SF network with $\lambda \approx
2$~\cite{Barrat_pnas}. We model the dynamic transport process of the WAN as a
SF network with correlated weights representing the number of
passengers~\cite{Barrat_pnas}. We study robustness and scaling of the optimal
paths and find good agreement with the real WAN results.

\section{Networks with topologically correlated weights}

\subsection{Network model with topologically correlated weights}

In studies on weighted networks such as the WAN and {\it E. Coli} metabolic
network, the weights on the links represent different quantities. The
weight in {\it E. Coli} metabolic network represents the flux of a link,
which represents the relative activity of the reaction on that
link~\cite{Almaas_ecoli}. For both real networks, the weight associated with
a link measures the preference level of that link.  The higher the weight is,
the easier it is to go through that link. In the sense, the inverse of the
weight on both networks is actually a plausible evaluation of the ``cost'' to
traverse that link. In the WAN, the ``cost'' includes the airfare, time and
convenience, etc. We assume that the higher is the cost on a link, the less
traffic it has.

From this perspective, we apply the optimization problem~\cite{Lidia_op_prl}
to the SF network with generalized correlated weight. To each link connecting
a pair of nodes $i$ and $j$ in a SF network we assign a weight $w_{ij}$,
representing the cost to transverse that link, with the form:
\begin{equation}
  w_{ij} \equiv x_{ij} (k_i k_j)^{\alpha},
  \label{weight_def}
\end{equation}
where $x_{ij}$ is a random number, $\alpha$ is a parameter that controls the
strength of the correlation between the topology and the weight and $k_i$,
$k_j$ is the degree of node $i$ and node $j$. The random number $x_{ij}$
could be chosen to mimic the statistical distribution of the real network and
in this paper $x_{ij}$ is taken from a uniform distribution between $0$ and
$1$\cite{FN_1}. Here we will be interested in the entire range of $\alpha$.

The minimum spanning tree (MST) of SF networks with correlation of
Eq.~(\ref{weight_def}) represents the structure carrying the maximum total
traffic in real WAN network and the skeleton of the most used paths in {\it
E. Coli} metabolic network. The MST is the tree spanning all nodes of the
network with the minimum total weight. With weights according to
Eq.~(\ref{weight_def}) and $\alpha = -0.5$, the MST is the same as the tree
that maximizes the total traffic out of all possible spanning trees of the
WAN network because the structure of the MST is only determined by the
relative order of the links according to the weight, which is preserved under
the inverse transformation. For this reason, in our simulation, we only show
results for $\alpha = 1, 0, -1$ representing $\alpha > 0$, $\alpha = 0$ and
$\alpha < 0$~\cite{Macdonald_ecoli, FN_alpha}. As an optimized tree, the MST
playing the role of the network skeleton, is widely used in different fields,
such as the design and operation of communication networks, the traveling
salesman problem, the protein interaction problem, optimal traffic flow and
economic networks~\cite{Khan_tech, Skiena_book_mst, Fredman_mst, Kruskal_mst,
Macdonald_ecoli, mst_eco_1, mst_eco_2}. Thus studying the effect of
correlations of the type of Eq.~(\ref{weight_def}) on the structure of the
MST may explain the transport on such weighted networks and possibly will
lead to better understanding of the origin of such correlations in real
networks. Moreover, the MST is the union of all the optimal paths in strong
disorder (SD) limit~\cite{Sameet_mst}, where one link on a path dominates the
total weight of the path. Thus, the scaling of the optimal paths also reveals
an important aspect of the structure of the MST.

\subsection{Scaling of the length of the optimal path}
\label{section_opt_path}

The case $\alpha = 0$ represents the uncorrelated case, which is studied in
Ref~\cite{Lidia_op_prl}. In the SD regime, the total weight of a path is
controlled by a single link with the highest cost on that path~\cite{Porto,
Cieplak:op1, Lidia_op_prl, Sameet_mst, Strelniker, zhenhua, tomer}. For
uncorrelated SF networks, the length of the optimal path in SD, $\ell_{\rm
opt}$ scales with $N$ as~\cite{Lidia_op_prl}:
\begin{eqnarray}
   \ell_{\rm opt} \sim \left
  \{\begin{array}{ll}
    N^{\nu_{\rm opt}} & ~~~\lambda > 3\\
    \ln^{\lambda - 1} N & ~~~2 < \lambda \le 3,
    \end{array}\right.
  \label{lopt_uncorrelated_sd_scaling}
\end{eqnarray}
with $\nu_{\rm opt} = 1/3$ for $\lambda \ge 4$ and $\nu_{\rm opt} = (\lambda -
3) / (\lambda -1)$ for $3 < \lambda < 4$.

From the definition of the weights (see Eq.~(\ref{weight_def})), in the case
$\alpha \neq 0$, we expect that correlations will affect the links connected
to the high degree nodes (hubs). The optimal paths behave either as ``{\it
hub-phobic}'' or as ``{\it hub-philic}'' depending on whether $\alpha$ is
positive or negative. {\it Hub-phobic} ($\alpha > 0$) means that the optimal
paths dislike to go through hubs because the cost is higher. {\it Hub-philic}
($\alpha < 0$) means that the optimal paths like to go through hubs because
they cost less. Thus $\alpha$ is a parameter controlling the importance level
of hubs in the optimized transport process. Due to the importance of the hubs
in SF networks, we expect that the scaling of $\ell_{\rm opt}$ will be
affected by such correlations. We expect in the hub-phobic regime, that the
optimal paths should be topologically similar to an ER
graph~\cite{Lidia_op_prl}, where there are almost no hubs. Thus $\ell_{\rm
opt}$ will scale with $N$ with exponent $1/3$ as in ER networks
~\cite{Lidia_op_prl} for all the values of $\lambda > 2$. On the other hand,
in the hub-philic regime, the hubs are preferred and the optimal paths will
avoid small or intermediate degrees, which are less important for the SF
networks and have minor influence on the optimal paths. Thus in this case we
expect that $\ell_{\rm opt}$ will scale similarly to the uncorrelated case.

The SD limit can be reached by approaching absolute zero temperature if
passing through a link is regarded as an activation process with a random
activation energy $\varepsilon_{ij}$ and $w_{ij} \equiv \exp(\beta
\varepsilon_{ij})$, where $\beta$ is the inverse temperature, since for
$\beta \to \infty$ a single link in the path dominates the sum $\sum w_{ij}$
along the path. For a general form of weight, which is not exponential but
rather such as Eq.~(\ref{weight_def}), the SD limit can be reached by
generating the MST because it provides the optimal path between any two nodes
of a network in the SD limit~\cite{Sameet_mst}. In this paper, we use the MST
to compute the optimal paths in the SD limit.

To model the WAN, we generate the SF networks using the Molloy-Reed algorithm
with the constraint of disallowing parallel links (two or more links
connecting the same two nodes) and self-loops (node connecting with
itself)~\cite{Molloy_Reed_book, Molloy_Reed}. Then for each link ${ij}$ in
the network, we assign a weight according to Eq.~(\ref{weight_def}).

We build the MST on the largest connected component of a graph using Prim's
algorithm~\cite{network_flow_book}. The tree starts from any node in the
largest connected component and grows to the nearest neighbor of the tree
with the minimum cost. This process ends when the MST includes all the nodes
of the largest connected component. This process is analogous to invasion
percolation used in physics~\cite{Bunde_book}. As the MST is a tree, there is
only a single path between any pair of nodes, which is the optimal path in
the SD limit~\cite{Sameet_mst}.

An equivalent algorithm to find the MST is the ``bombing
algorithm''~\cite{Cieplak:op1, Lidia_op_prl}, where we start with the full
network and remove links in descending order of the weight only when the
removal of a link does not disconnect the graph. The algorithm ends, and the
MST is obtained, when no more links that can be removed without disconnecting
the graph. Then we calculate the mean value of $\ell_{\rm opt}$ between a
pair of randomly chosen nodes and average over many pairs and many
realizations.

\begin{figure}
  \includegraphics[width = \textwidth]{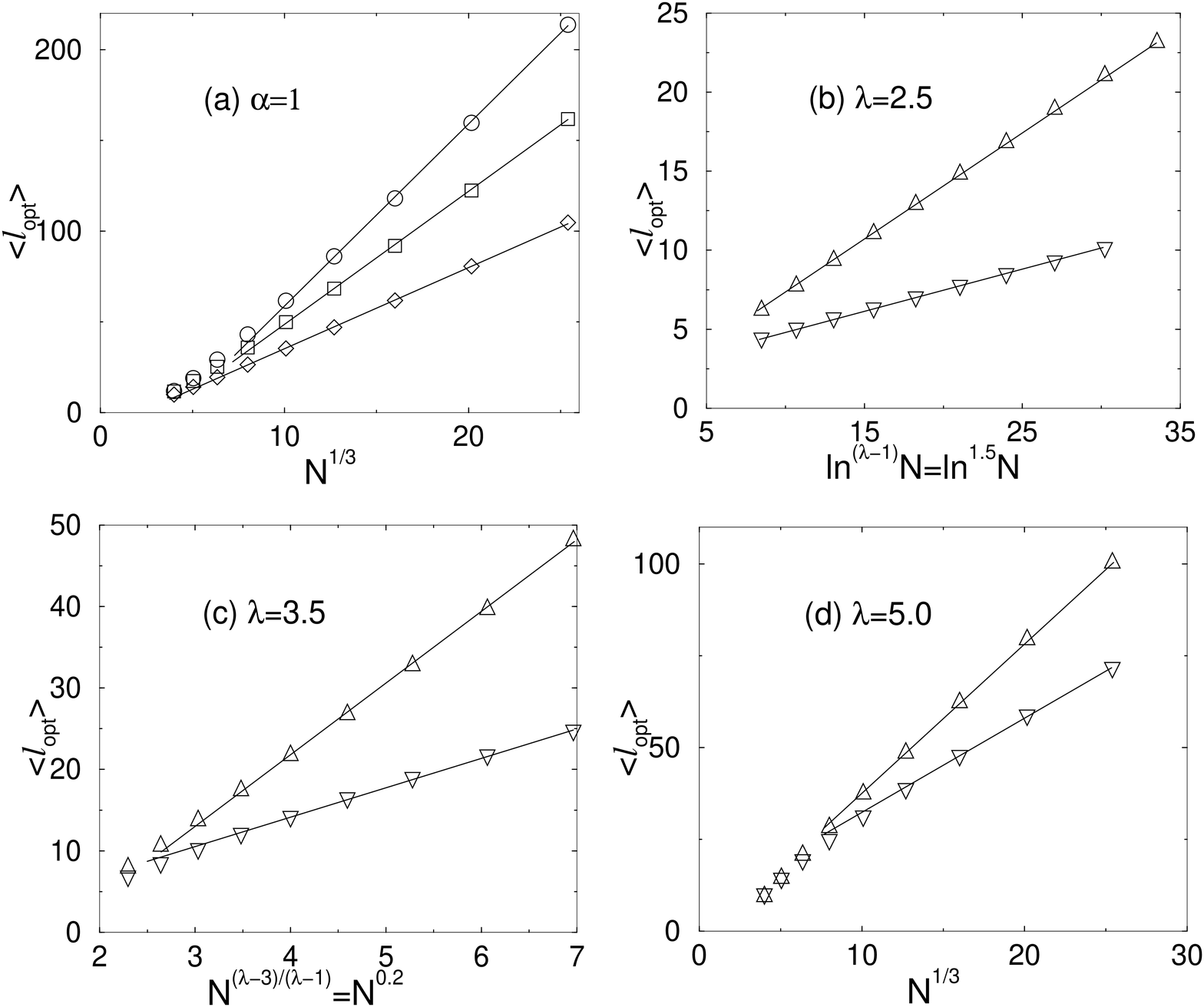}
  \caption{$\langle \ell_{\rm opt} \rangle$ as a function of (a) $N^{1/3}$,
    (b) $\ln^{(\lambda - 1)} N$, (c) $N^{(\lambda - 3)/(\lambda - 1)}$ and
    (d) $N^{1/3}$ in SD for (a) $\alpha = 1$ with: $\lambda = 5$
    ($\bigcirc$), $\lambda =3.5$ ($\Box$) and $\lambda =2.5$ ($\Diamond$);
    (b) $\lambda = 2.5$ with: $\alpha = 0$ ($\bigtriangleup$) and $\alpha =
    -1$ ($\bigtriangledown$); (c) $\lambda = 3.5$ with: $\alpha = 0$
    ($\bigtriangleup$) and $\alpha = -1$ ($\bigtriangledown$) and (d)
    $\lambda = 5.0$ with: $\alpha = 0$ ($\bigtriangleup$) and $\alpha = -1$
    ($\bigtriangledown$).}
  \label{sf_cwr_all_alpha_extra}
\end{figure}

Figure~\ref{sf_cwr_all_alpha_extra} (a) shows $\ell_{\rm opt}$ as a function
of $N^{1/3}$ for several values of $\lambda > 2$ with $\alpha = 1$. The
linear behavior supports the expected scaling for the SF networks in the
hub-phobic regime ($\alpha > 0$):
\begin{equation}
  \ell_{\rm opt} \sim N^{1/3} \qquad \qquad \qquad [\lambda > 2, \alpha >0].
  \label{lopt_scaling_alpha_gt_0}
\end{equation}
Eq.~(\ref{lopt_scaling_alpha_gt_0}) supports our hypothesis that if the hubs
are dynamically avoided, the SF networks behave the same as an ER networks.

In the hub-philic regime ($\alpha < 0$) , we find that $\ell_{\rm opt}$
scales the same as for the uncorrelated case ($\alpha = 0$). In
Figs.~\ref{sf_cwr_all_alpha_extra} (b), (c) and (d), we plot the scaling
behavior of $\ell_{\rm opt}$ as a function of $N$ according to the scaling
found for the uncorrelated case~\cite{Lidia_op_prl}. Our numerical results
suggest that the scaling behavior of $\ell_{\rm opt}$ in the hub-philic
regime has the same scaling form of Eq.~(\ref{lopt_uncorrelated_sd_scaling})
as the uncorrelated case. However, in the hub-philic regime, $\ell_{\rm opt}$
is much smaller than that of the uncorrelated case, which shows the advantage
for the optimal paths to pass through the hubs.

\subsection{Robustness}
\label{section_robust}
Percolation properties of random networks with given degree distributions
have been extensively studied \cite{Cohen_robust, Callaway, Newman_random,
Cohen_intentional_attack}. One of the most striking
results~\cite{Cohen_robust}, due to its important implications, is the
absence of a percolation threshold in the uncorrelated random SF networks
with $\lambda < 3$. In other words, in this type of network, one has to
remove almost all nodes before the network collapses into disconnected
components because the few hubs always keep the remaining network
connected~\cite{Cohen_robust}. Translated into the epidemic context, it means
that an epidemic threshold below which the epidemics cannot propagate
approaches zero as $N \to \infty$.

Robustness of a network can be characterized by the fraction of the links or
nodes one has to remove in order to disconnect the whole
network. Nevertheless, almost all the analytical results on robustness
obtained up to now implicitly refer to uncorrelated networks and little is
known about the effects of topologically correlated weights on the
percolation properties of networks. To test the robustness in weighted
correlated networks, we calculate the ratio $P_\infty \equiv \langle S
\rangle /N $ as a function of $q$, where $\langle S \rangle$ is the average
mass of the largest remaining cluster, $N$ is the mass of the whole network
and $q$ is the fraction of removed links. $P_\infty$ basically is the
probability to find a node belonging to the giant component. In order to
compute $P_\infty$ after building the network, we assign weights on links
according to Eq.~(\ref{weight_def}). Then we remove a fraction $q$ of links
in descending order of weights and calculate $P_\infty$ through $\langle S
\rangle / N$. In order to identify the percolation threshold $p_c=1-q_c$ we
compute also the average mass of the second largest component $\langle S_{\rm
2} \rangle$, and estimate $q_c$ from its maximum value \cite{Bunde_book}.

\begin{figure}
  \includegraphics[width=0.37\textwidth, angle = -90]{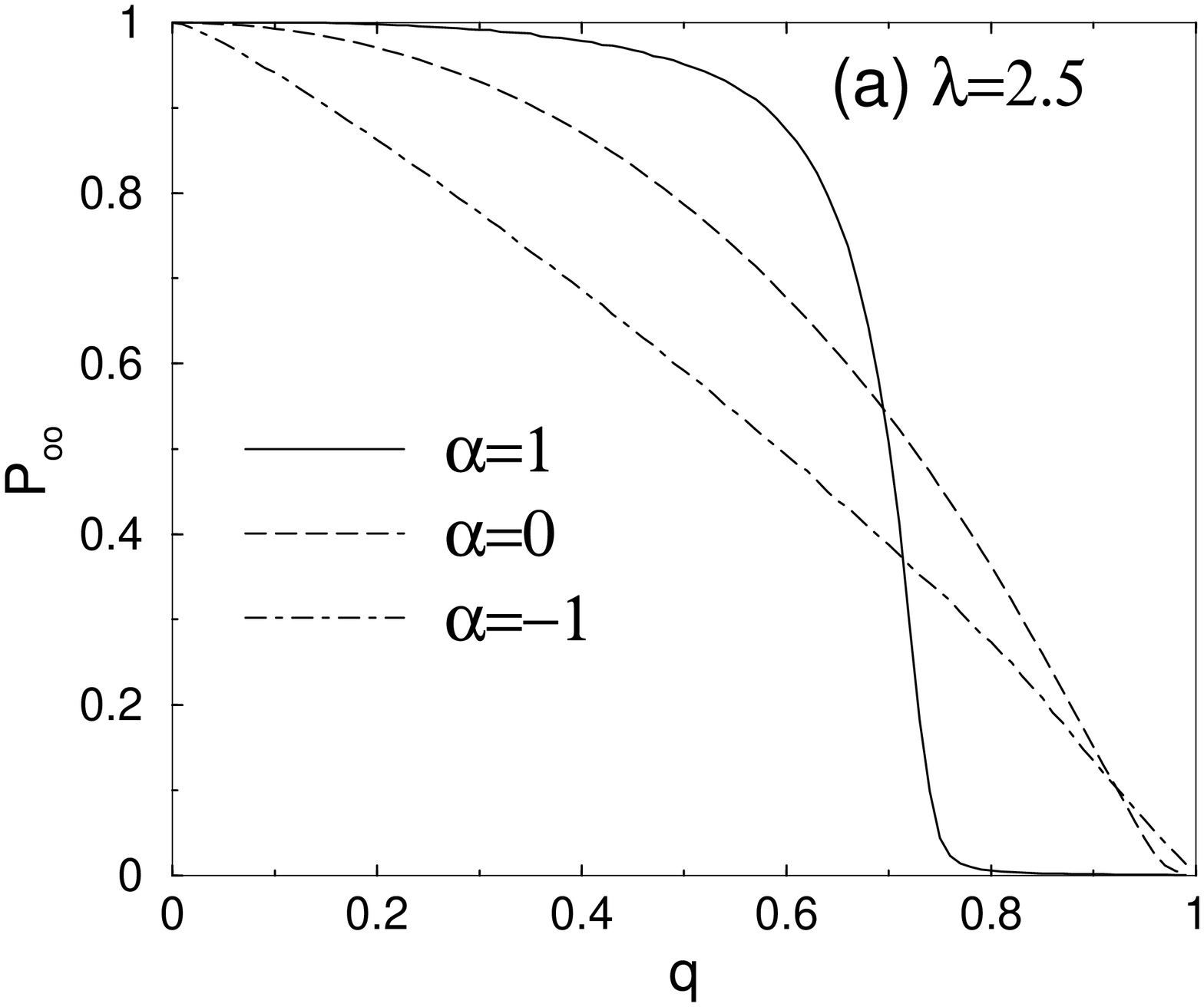}
  \includegraphics[width=0.37\textwidth, angle = -90]{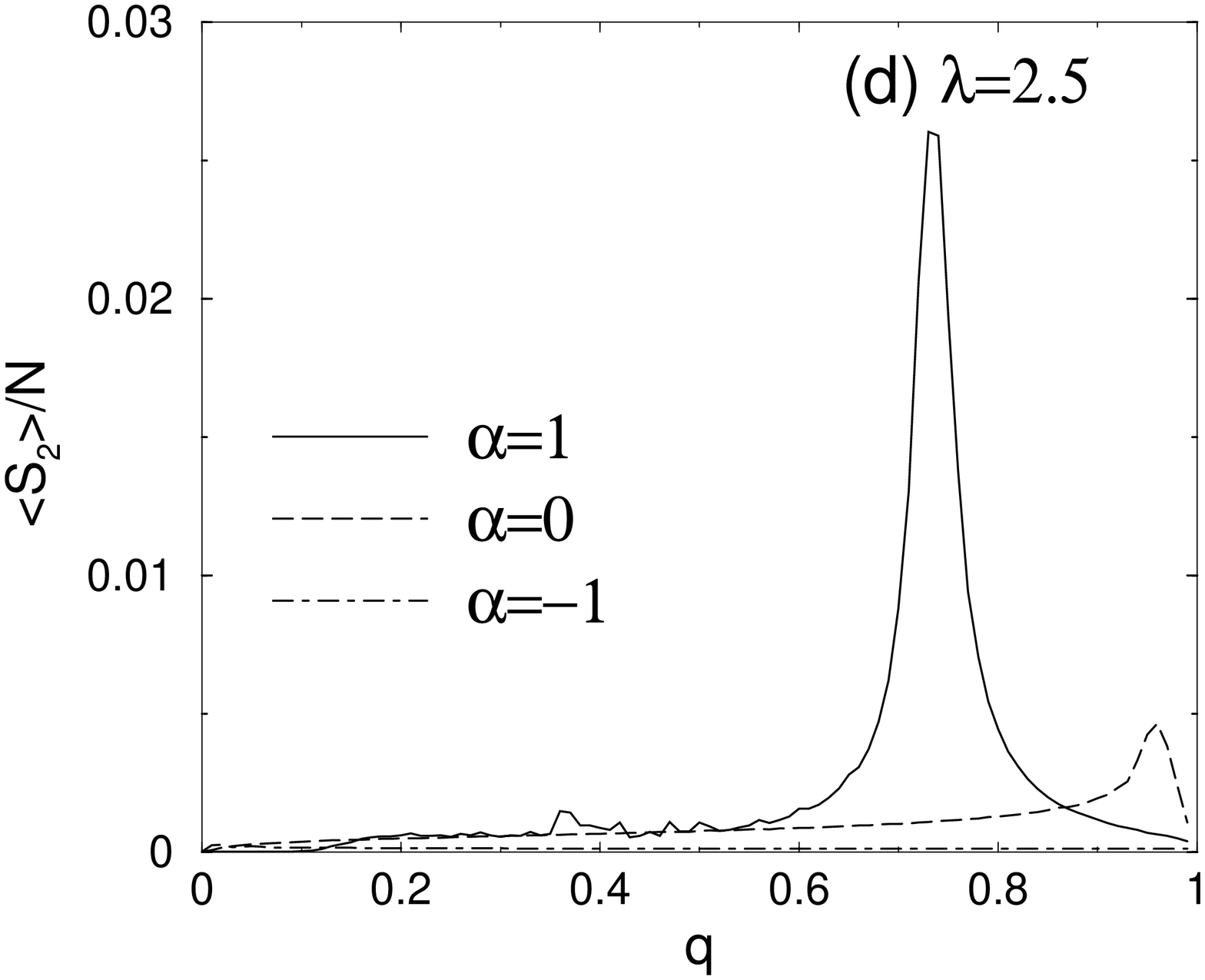}
  \includegraphics[width=0.37\textwidth, angle = -90]{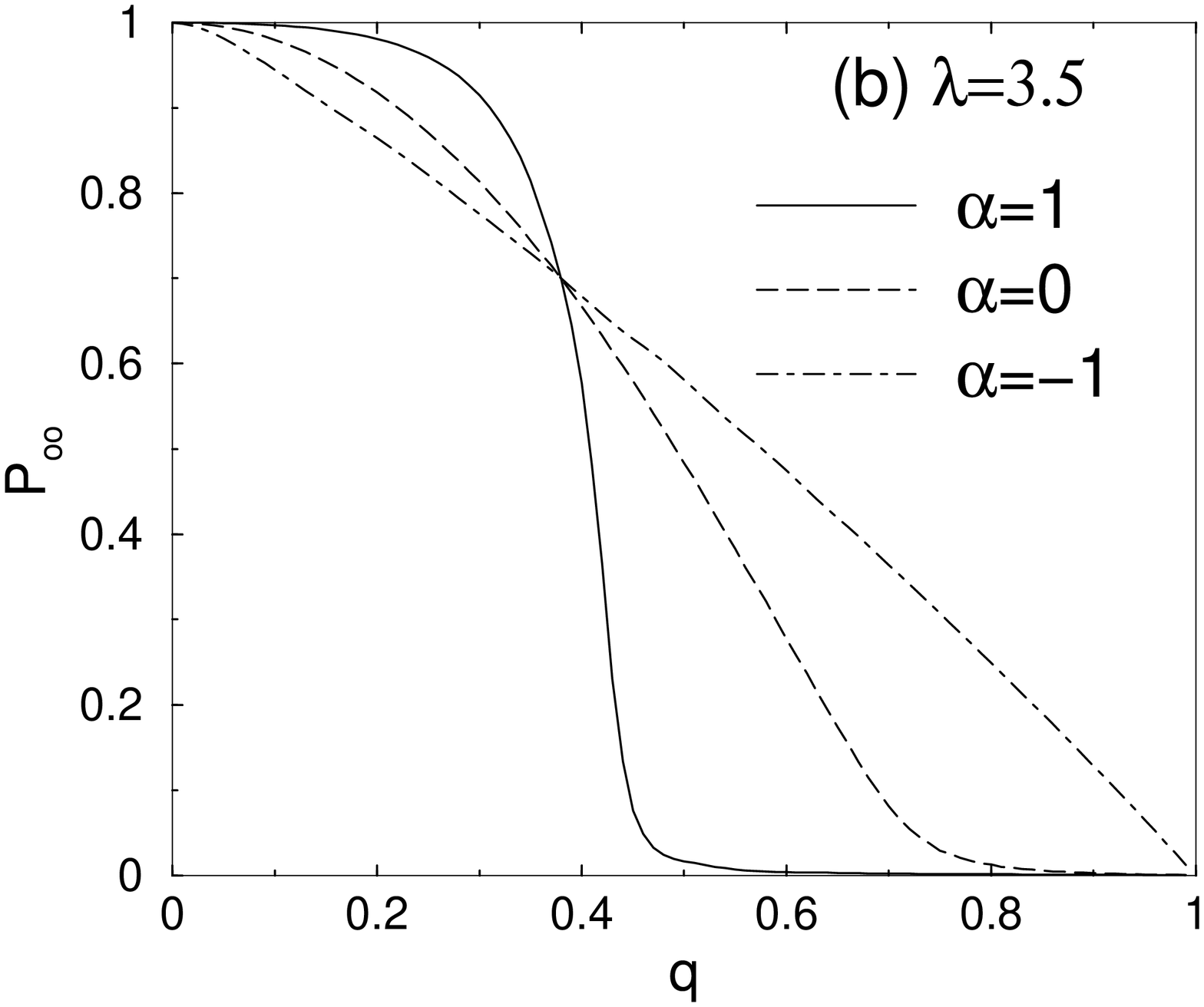}
  \includegraphics[width=0.37\textwidth, angle = -90]{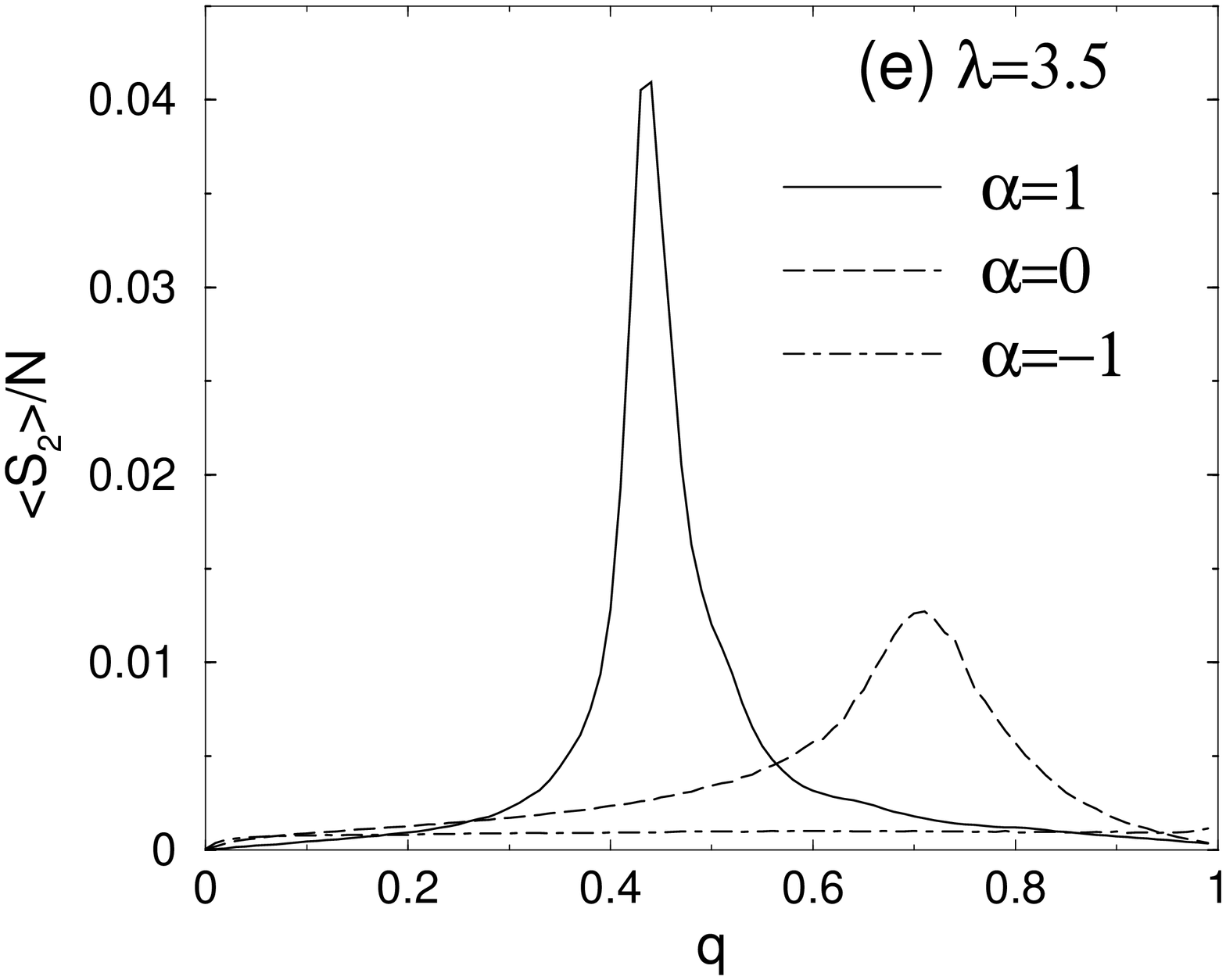}
  \includegraphics[width=0.37\textwidth, angle = -90]{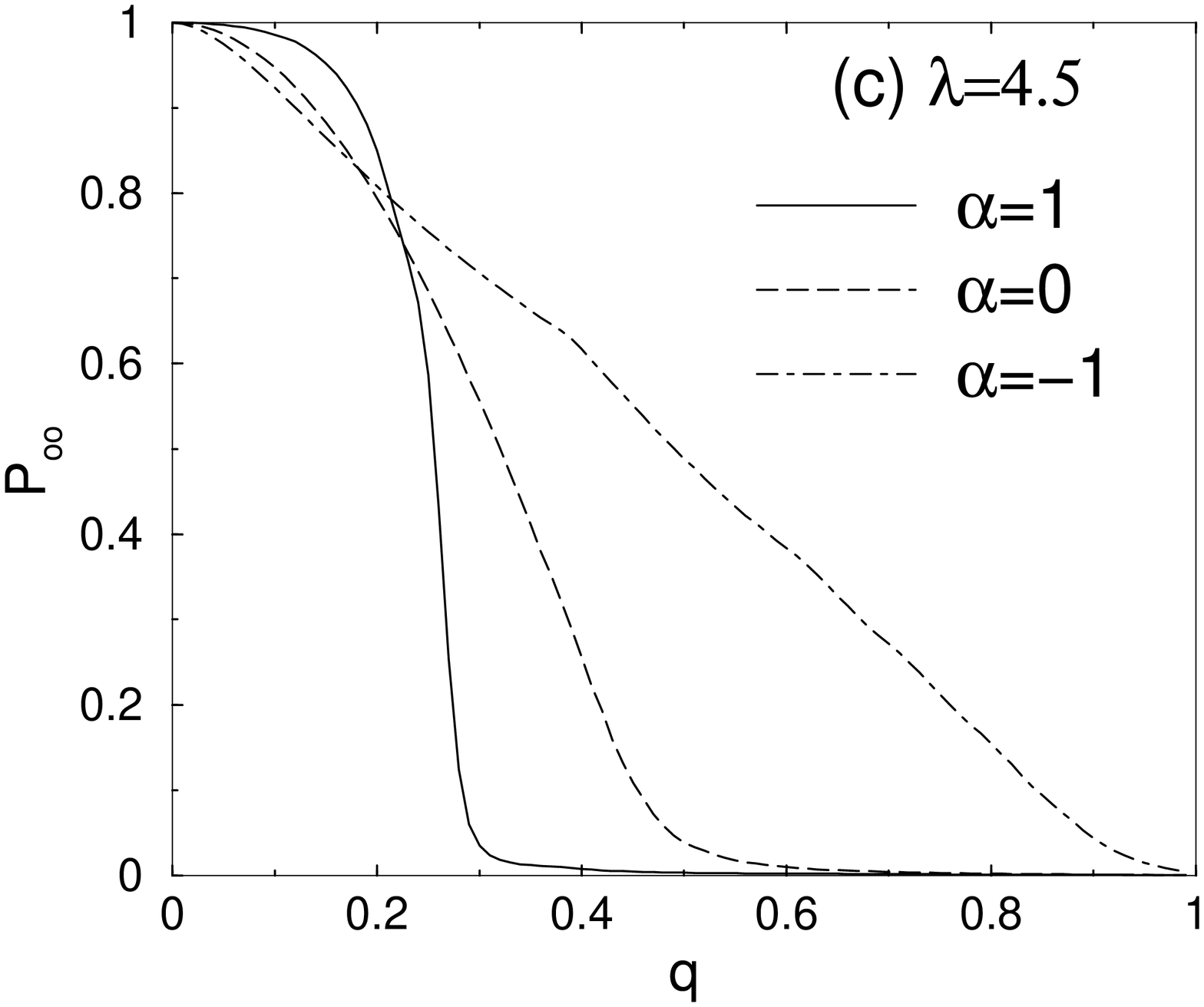}
  \includegraphics[width=0.37\textwidth, angle = -90]{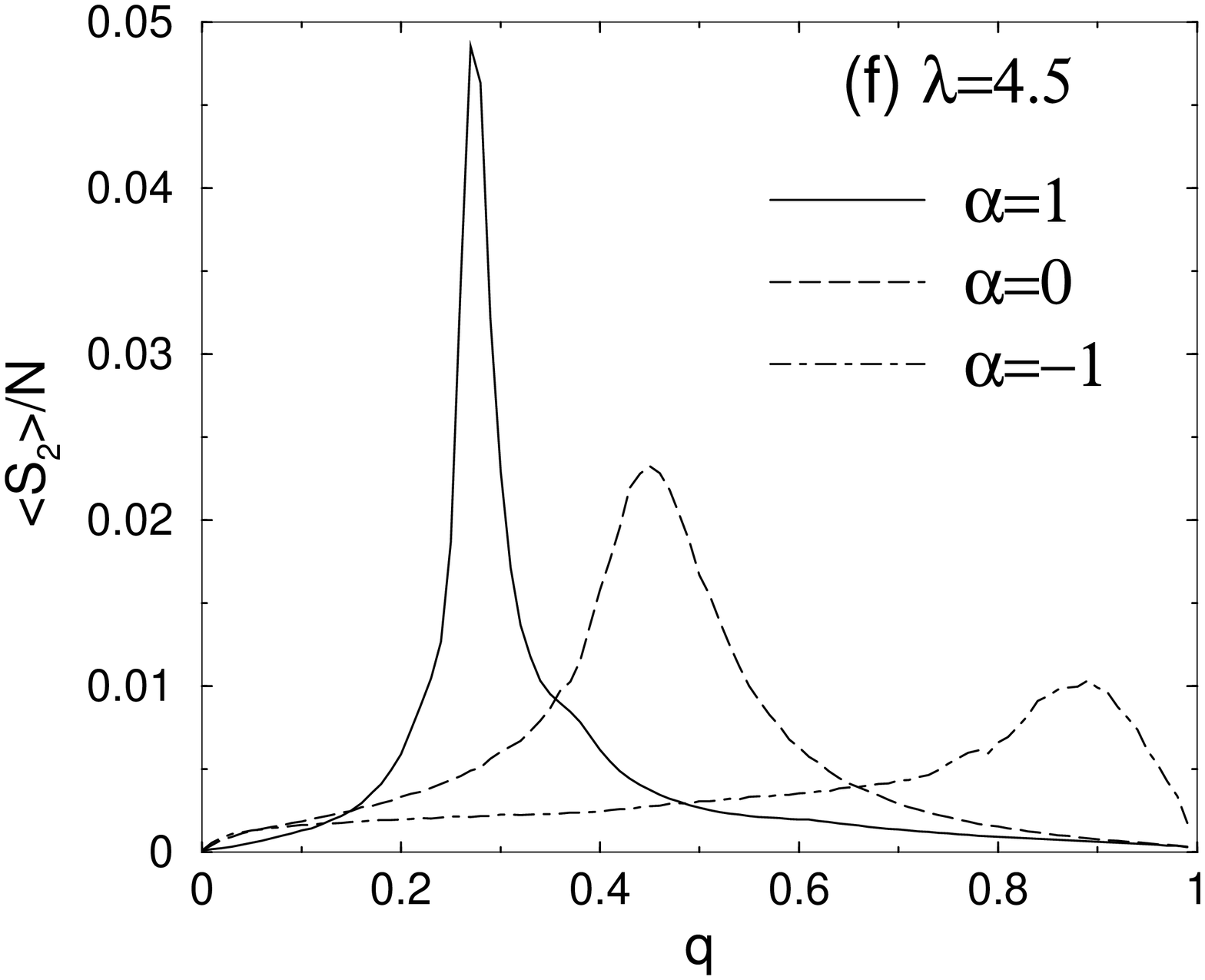}
  \caption{Left column (a)-(c): $P_{\infty}$ as a function of $q$, the
    concentration of removed links with (a) $\lambda = 2.5$, (b) $\lambda =
    3.5$ and (c) $\lambda = 4.5$. Right column (d)-(f): $\langle S_{2}
    \rangle / N$ as a function of $q$ with (d) $\lambda = 2.5$, (e) $\lambda
    = 3.5$ and (f) $\lambda = 4.5$.}
  \label{Robust}
\end{figure}

In Figure~\ref{Robust} we show the results for networks with $N = 8192$ and
different values of $\lambda$ and $\alpha$. The peak of $\langle S_{\rm 2}
\rangle$ indicates the position of $p_c(N)$. We can see that for uncorrelated
case ($\alpha = 0$), as expected $p_c(N)$ is close to zero for $\lambda =
2.5$. In the hub-phobic ($\alpha > 0$) regime, $p_c(N)$ is always finite as
expected due to its topological similarity with ER networks. However, in the
hub-philic regime ($\alpha < 0$), firstly we observe that the $p_c(N)$ is
always smaller than uncorrelated case, which means that networks with $\alpha
< 0$ are more robust than uncorrelated ones. Secondly, $p_c(N=8192)$ in the
hub-philic regime ($\alpha < 0$) is very close to zero even for $\lambda =
3.5$, which suggests that $p_c(\infty)$ might be zero. This is very unusual
because it is known that in the uncorrelated case, $p_c(\infty)$ is finite
for $\lambda > 3$ \cite{Cohen_robust}. A similar behavior $p_c(N)$ close to
zero is obtained for $\alpha < 0$ and $3 < \lambda < 4$. Thus, for these
values of $\lambda$, the case of the hub-philic regime might correspond to a
new universality class. This question will be discussed in
Sec.~\ref{section_analytical_method}.

\begin{figure}
  \includegraphics[width=0.37\textwidth, angle = -90]{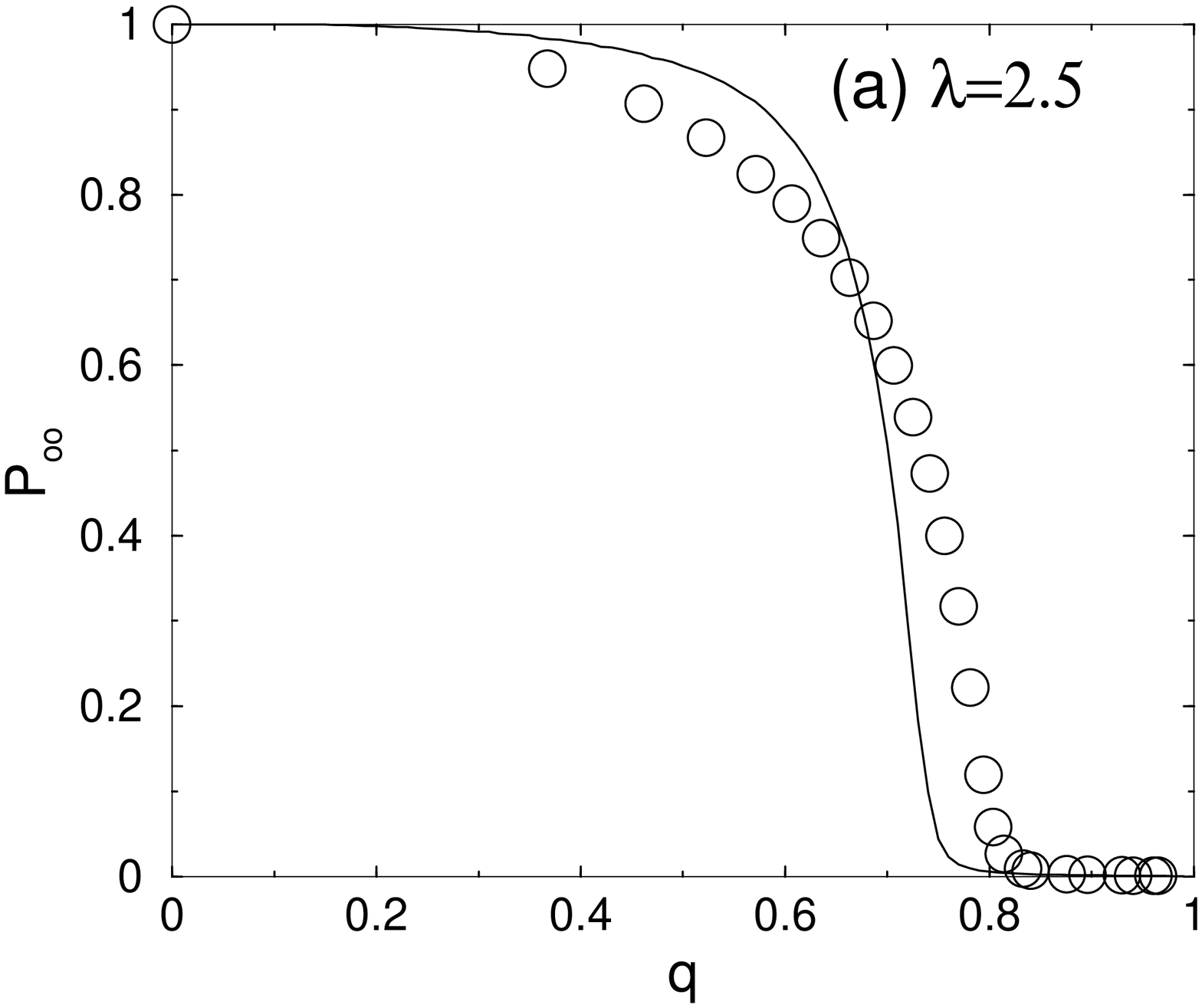}
  \includegraphics[width=0.37\textwidth, angle = -90]{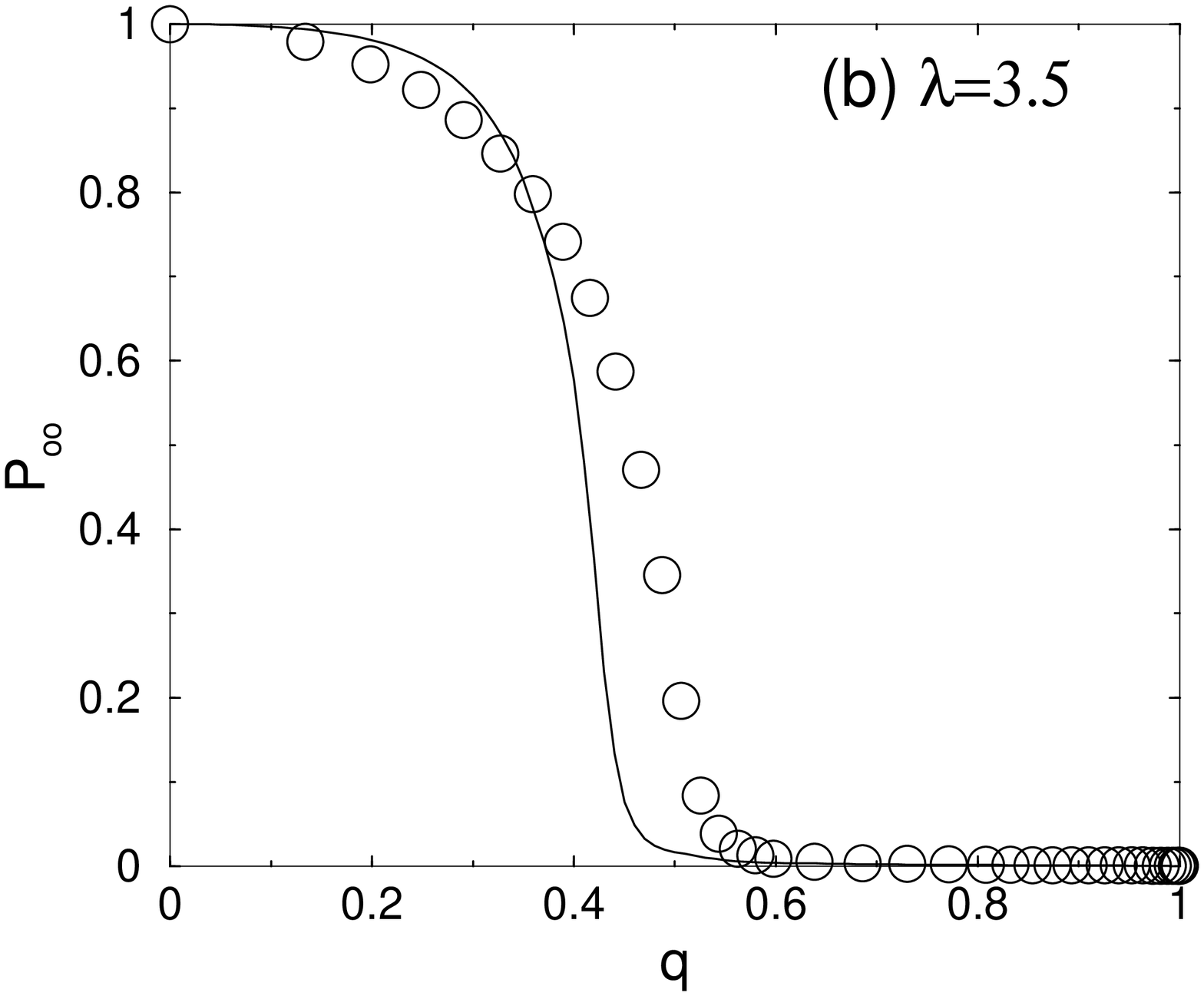}
  \includegraphics[width=0.37\textwidth, angle = -90]{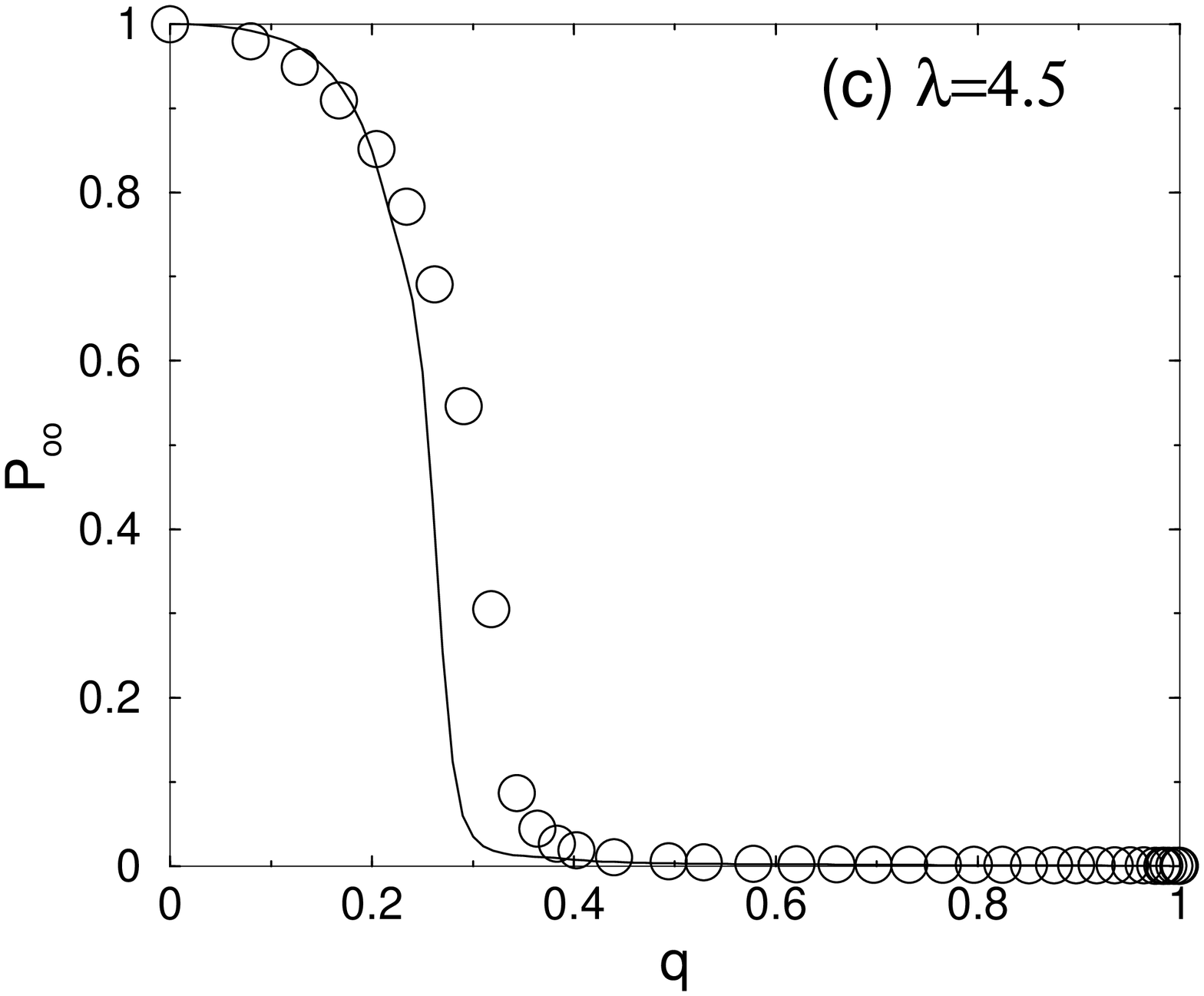}
  \caption{$P_{\infty}$ as a function of $q$ with (a) $\lambda = 2.5$, (b)
    $\lambda = 3.5$ and (c) $\lambda = 4.5$ for {\it Strategy I}
    ($\bigcirc$): attacking nodes according to degree of nodes and {\it
    Strategy II} (solid line): attacking links according to its weight for
    the SF networks in the hub-phobic regime ($\alpha > 0$).}
  \label{attack_strategy}
\end{figure}

Robustness is directly related to the attacking and immunization
strategies. The calculations in Fig.~\ref{Robust} show the robustness of
networks against attacks according to the link weights. There are other types
of non-random attacking strategies such as: ({\it Strategy I}): intentionally
attacking node according to its degree~\cite{Cohen_intentional_attack,
Gallos_netattack}; ({\it Strategy II}): intentionally attacking links
according to its weight in the SF network at the hub-phobic regime ($\alpha >
0$). It was found that {\it Strategy I} is an efficient way of attacking a SF
network~\cite{Cohen_intentional_attack, Gallos_netattack}. For {\it strategy
II}, in the hub-phobic regime, the links connecting to hubs will be attacked
first. Next we compare the efficiency of these two types of intentional
attack strategies. To fairly compare the two strategies, we calculate
$P_\infty$ of both intentional attack strategies as a function of $q$, the
fraction of removed links. For {\it Strategy I}, attacking a node is
equivalent to attack all the links of that node. In
Fig.~\ref{attack_strategy}, we see that these two strategies are similar to
each other. Comparing the position of $q_c$ for these two strategies, we can
conclude that {\it Strategy II} is actually slightly more efficient than {\it
Strategy I}.

From Fig.~\ref{Robust} (a) we can see that
\begin{equation}
  p_c(\alpha = -1) < p_c(\alpha = 0) < p_c(\alpha = 1),
\end{equation}
a result which is not surprising because as $\alpha$ increases more central
links connecting high degree nodes are removed. A surprising result is that
\begin{equation}
  P_\infty(\alpha = 1) > P_\infty(\alpha = 0) > P_\infty(\alpha = -1)
\end{equation}
for a wide range of $q$ values, e.g., for $\lambda = 2.5$ below $q_c \approx
0.7$. This is probably due to the fact that when removing central links
connecting high degree nodes the remaining network does not become
disconnected as shown in the study of the k-core of
networks~\cite{Carmi_condmat}. The effect of removing those central links is
mostly to increase the distance.

\begin{figure}
  \includegraphics[width=0.6\textwidth, angle = -90]{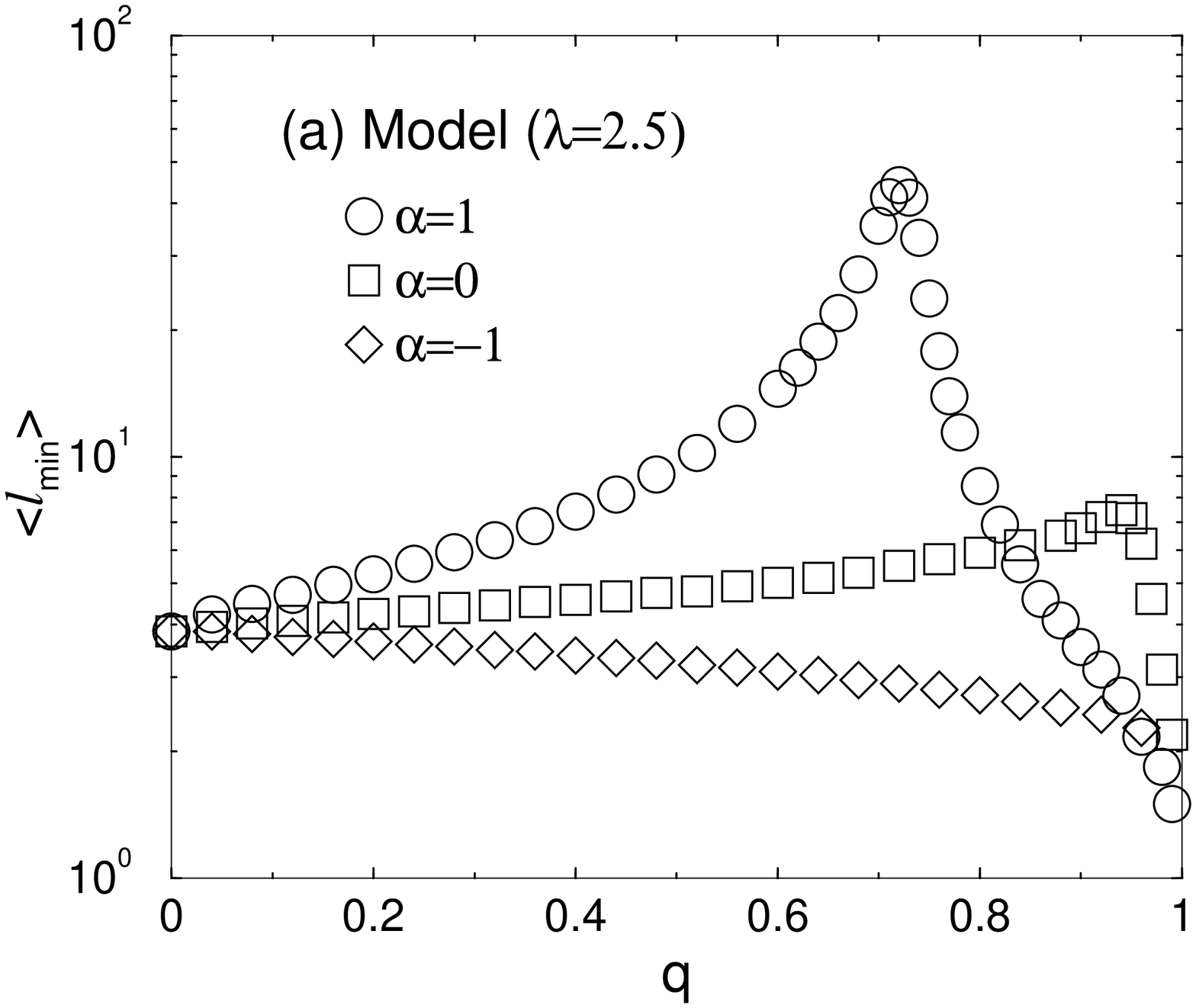}
  \includegraphics[width=0.6\textwidth, angle = -90]{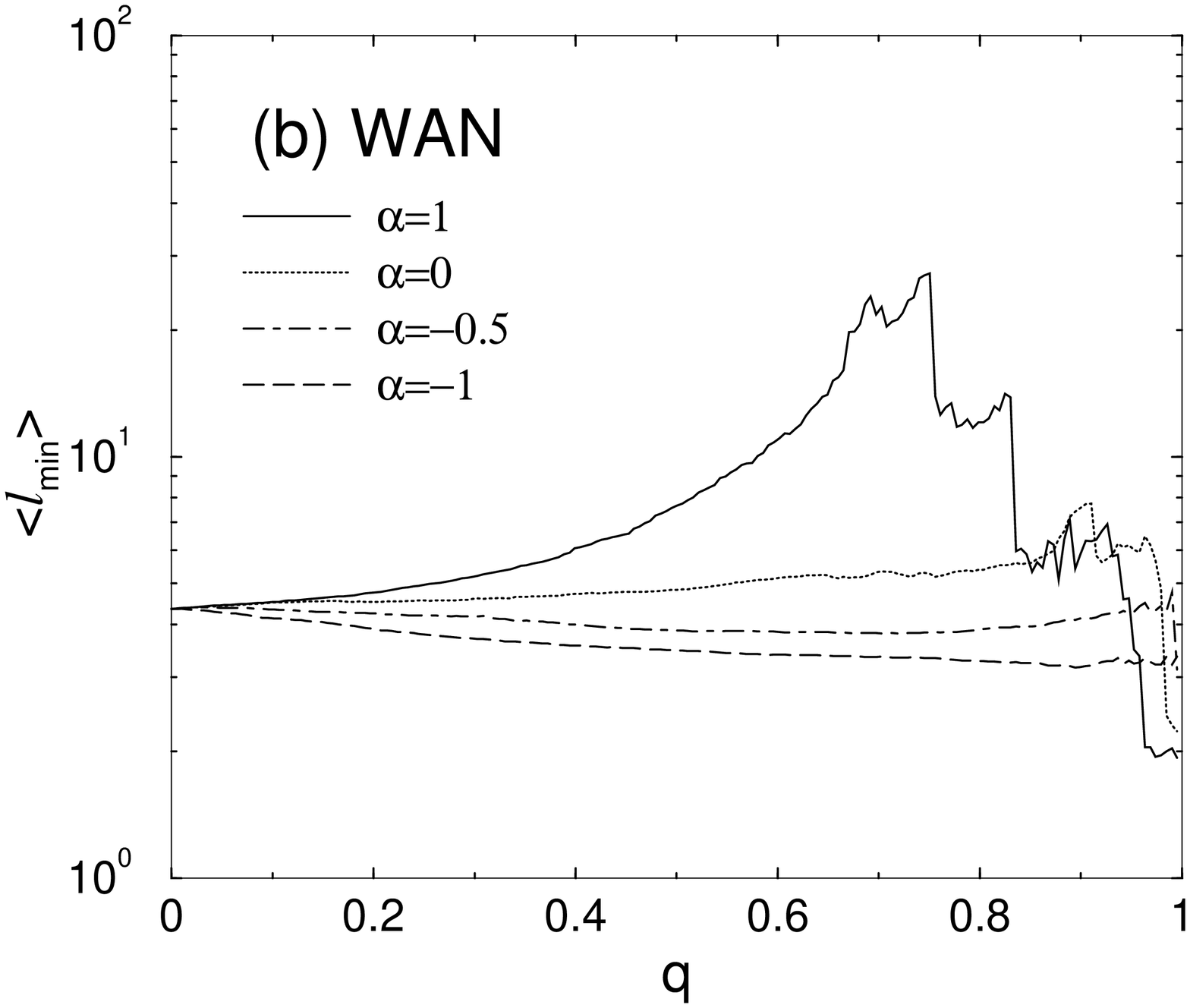}
  \caption{(a) $\langle \ell_{\rm min} \rangle$, the average length of the
    shortest path as a function of $q$, the concentration of removed links in
    descending order of the weight for simulated SF networks with $\alpha =
    1$ ($\bigcirc$), $\alpha = 0$ ($\Box$) and $\alpha = -1$
    ($\Diamond$). (b) The same calculation for the real WAN network with
    $\alpha = 1$ (solid line), $\alpha = 0$ (dotted line), $\alpha = -0.5$
    (dot-dashed line) and $\alpha = -1$ (dashed line). }
  \label{ave_dis_garph}
\end{figure}

To test our hypothesis we calculate the average length of the shortest path
$\ell_{\rm min}$ of the largest cluster as a function of $q$ for the $3$
different correlation cases. In Fig.~\ref{ave_dis_garph} (a) we plot $\langle
\ell_{\rm min} \rangle$ as function of $q$.  We can see that
\begin{equation}
  \langle \ell_{\rm min} \rangle_{\alpha = 1} > \langle \ell_{\rm min}
  \rangle_{\alpha = 0} > \langle \ell_{\rm min} \rangle_{\alpha = -1}
\end{equation}
for a wide range of $q$ as predicted. The peaks for $\alpha = 1$ and $\alpha
= 0$ are at the same position as the corresponding $q_c$ of Fig.~\ref{Robust}
(d) because when $q > q_c$ the whole network is fragmented. We compute
$\langle \ell_{\rm min} \rangle$ also for the real WAN network and obtain
similar results (See Fig.~\ref{ave_dis_garph} (b) and
Sec.~\ref{sec_WAN_SIM_comp}).

\subsection{Percolation analysis of correlated scale-free networks}
\label{section_analytical_method}

An important question is whether for $\alpha < 0$ and $3 < \lambda < 4$,
$p_c(\infty)$ is zero. In simulations we observe that $p_c(\infty)$ is
incredibly close to zero for $3 < \lambda < 4$. However, it might be a result
of finite size effects and $p_c$ is indeed very close to zero but not
zero. To further test this possibility, we propose the following analytical
method.

Near $p_c$, the structure of the giant component is like a tree.  The giant
component at the percolation threshold can be viewed as a growing process of
a tree.  Assuming that at some layer, the number of nodes with degree $k$ is
$n(k)$, then the number nodes of degree $k'$ at the next level $n(k')$
satisfies
\begin{equation}
  n(k') = \sum_k{n(k) (k - 1) \frac{k' {\cal P}(k')}{\langle k' \rangle}
    \phi(k', k)},
  \label{recursive_eq}
\end{equation}
where $\phi(k', k)$ is the probability that a node with degree $k$ is
connected to a node of degree $k'$. Here we are interested in the hub-philic
regime ($\alpha = -1$). Since we remove the links in descending order, we can
assume that any link with weight above $1/c$ are cut. The links left are only
links with weights below $1/c$, which is represented by the condition $k k' >
c$. Eq.~(\ref{recursive_eq}) can be therefore simplified to
\begin{equation}
  n(k') = \sum_{k = c/k'}^{\infty}{n(k) (k - 1) \frac{k' {\cal
	P}(k')}{\langle k' \rangle}}.
  \label{recursive_eq_final}
\end{equation}
The dimension of the vector $n(k)$ is actually the maximum degree $k_{\rm
max}$. For a SF network with system size $N$, $k_{\rm max} \sim N^{1/(\lambda
- 1)}$~\cite{Cohen_robust, Newman_review}. Thus controlling, $k_{\rm max}$,
the dimension of the vector $n(k)$ is actually equivalent to controlling the
system size $N$. For a fixed $k_{\rm max}$, if Eq.~(\ref{recursive_eq_final})
has at least one eigenvalue that is above $1$, the giant component will grow
to infinity, which means it is above $p_c$; on the other hand, if all of the
eigenvalues are below $1$, the system is below $p_c$. We basically change the
$c$ value and compute the normal of vector $n(k)$ recursively, and find the
critical $c$ value, $c^*$, at which the normal of vector $n(k)$ changes from
converging at $c^*$ to diverging at $c^* + 1$. Thus, from the definition of
percolation, $p_c(N)$ can be numerically calculated using the relation
\begin{equation}
  p_c(N) = \int_{k k' > c^*} \frac{k k' {\cal P}(k) {\cal
      P}(k')}{\langle k \rangle^2} dk dk'.
  \label{pc_analytic}
\end{equation}
If $c^*(k_{\rm max})$ diverges as $k_{\rm max} \to \infty$, from
Eq.~(\ref{pc_analytic}) follows that $p_c(\infty)$ is zero. Otherwise,
$p_c(\infty)$ is finite. Thus, we convert the question of whether
$p_c(\infty)$ is zero for $3 < \lambda < 4$ into the question of whether
$c^*(k_{\rm max})$ diverges with $k_{\rm max}$.

\begin{figure}
  \includegraphics[width=0.6\textwidth, angle = -90]{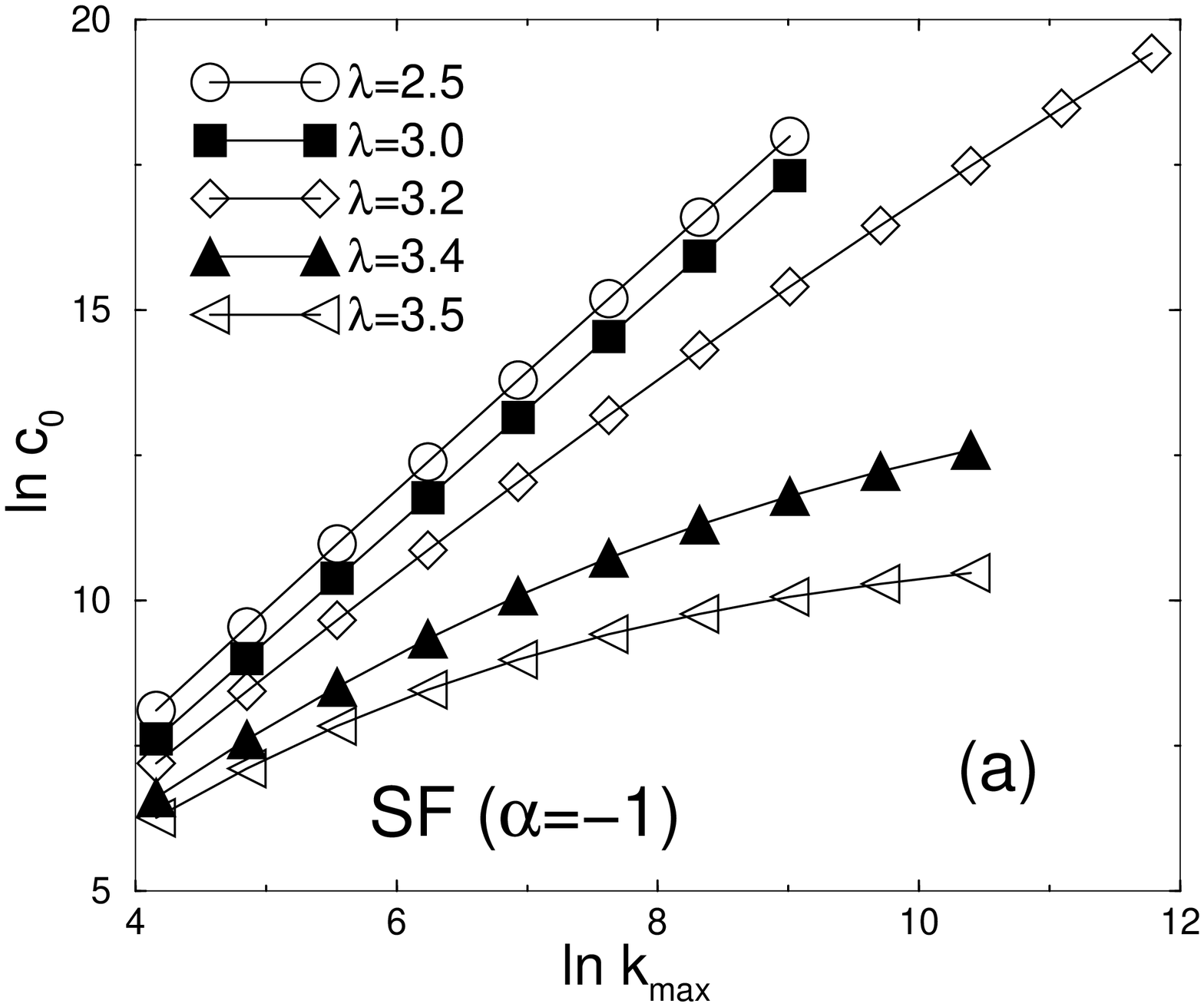}
  \includegraphics[width=0.6\textwidth, angle = -90]{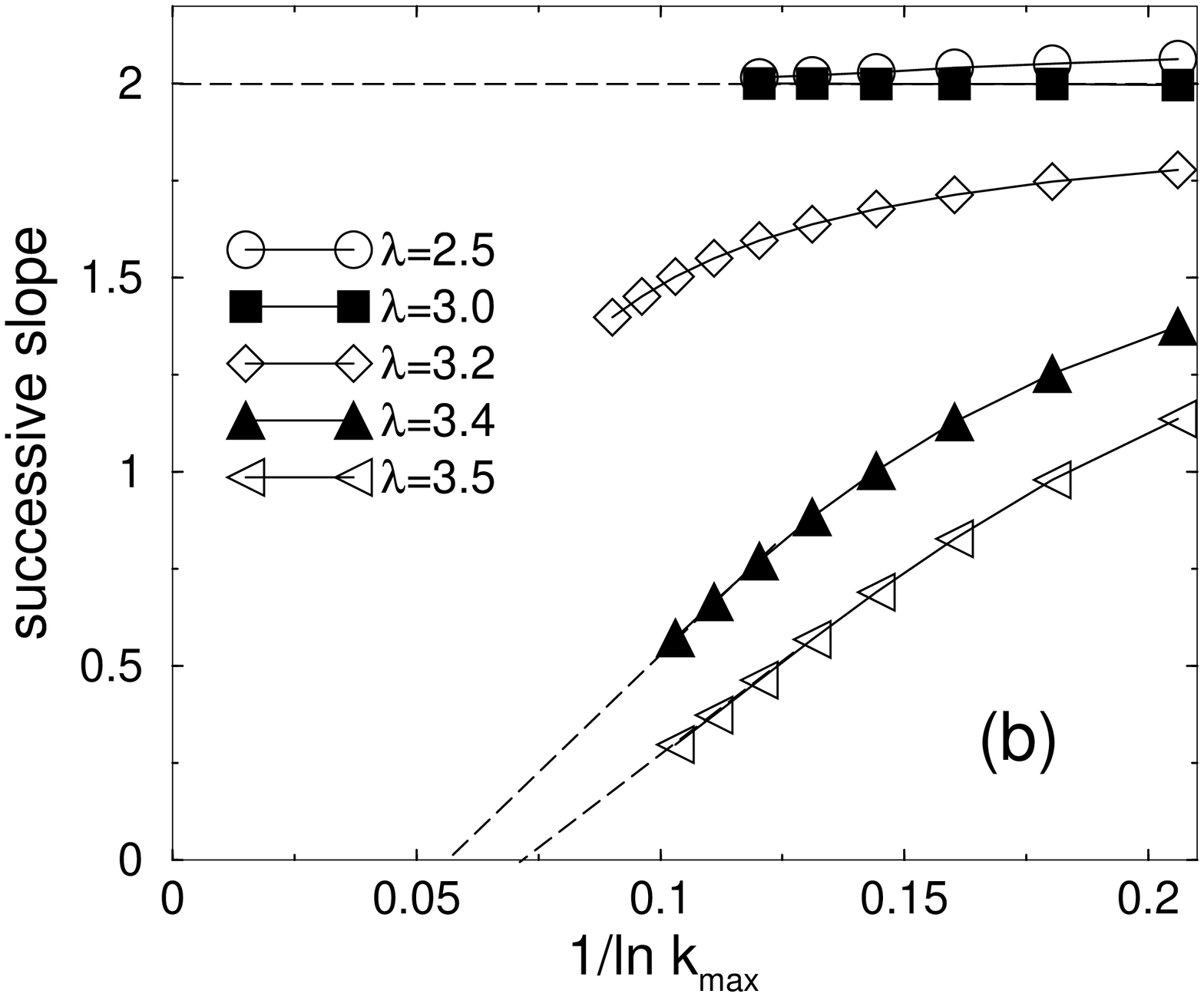}
  \caption{(a) $\ln c_0$ as a function of $\ln k_{\rm max}$. (b) Plot of
    the successive slopes of (a) as a function of $1 / \ln k_{\rm max}$. }
  \label{c_kmax}
\end{figure}

Figure~\ref{c_kmax} shows numerical results of $c^*(k_{\rm max})$ as $k_{\rm
max}$ increases. It shows that for $\lambda < 3$, $c^*$ grows with $k_{\rm
max}$ as a power law, which confirms that when $\lambda <3$, $p_c(\infty)$ is
zero~\cite{Cohen_robust}. However, for $3 < \lambda < 4$,
Fig.~\ref{c_kmax}(b) suggests that the successive slopes of $\ln c^*$ versus
$\ln k_{\rm max}$ approach zero for $3 < \lambda < 4$, which indicates a
finite $p_c(\infty)$ when $3 < \lambda < 4$. We can see the strong finite
size effect from Fig.~\ref{c_kmax} , where the convergence of $c^*$ for
$\lambda = 3.5$ happens only for $k_{\rm max} > 10^4$ corresponding to $N
\sim k_{max}^{\lambda -1} > 10^{10}$, which we are not able to reach in
simulations with present computing power.

\section{The impact of the Hub-philic correlation on real world networks}
\label{sec_WAN_SIM_comp}

Real world networks such as the WAN and the {\it E. Coli} metabolic network
are found to have a hub-philic type correlation ($\alpha =
-0.5$)~\cite{Barrat_pnas, Almaas_ecoli, Macdonald_ecoli}. Using WAN as an
example, the passengers tend to go through the large airports, which actually
shortens and optimizes the transport performance of the entire WAN
network. Our simulations show the advantage of networks in the hub-philic
regime over uncorrelated networks from the perspective of the optimal paths
in the SD regime (see Fig.~\ref{sf_cwr_all_alpha_extra} (b)-(d)). Also they
show that networks in the hub-philic regime are also more robust than
uncorrelated networks (see Fig.~\ref{Robust}). Real WAN have only one value
of $\alpha = -0.5$, so to see the effect of different $\alpha$ on the real
WAN network, we use the following method. Assigning the weights according to
Eq.~(\ref{weight_def}), we first compute $x^{\rm real}_{ij}$, the value of
$x_{ij}$ from the real weight of WAN. Thus we leave $\alpha$ as a parameter
and we can change its value to get either the hub-philic or the hub-phobic
regime. In this case, when $\alpha = -0.5$, the weight is the same as in the
original WAN network.

\begin{figure}
  \includegraphics[width=0.37\textwidth, angle = -90]{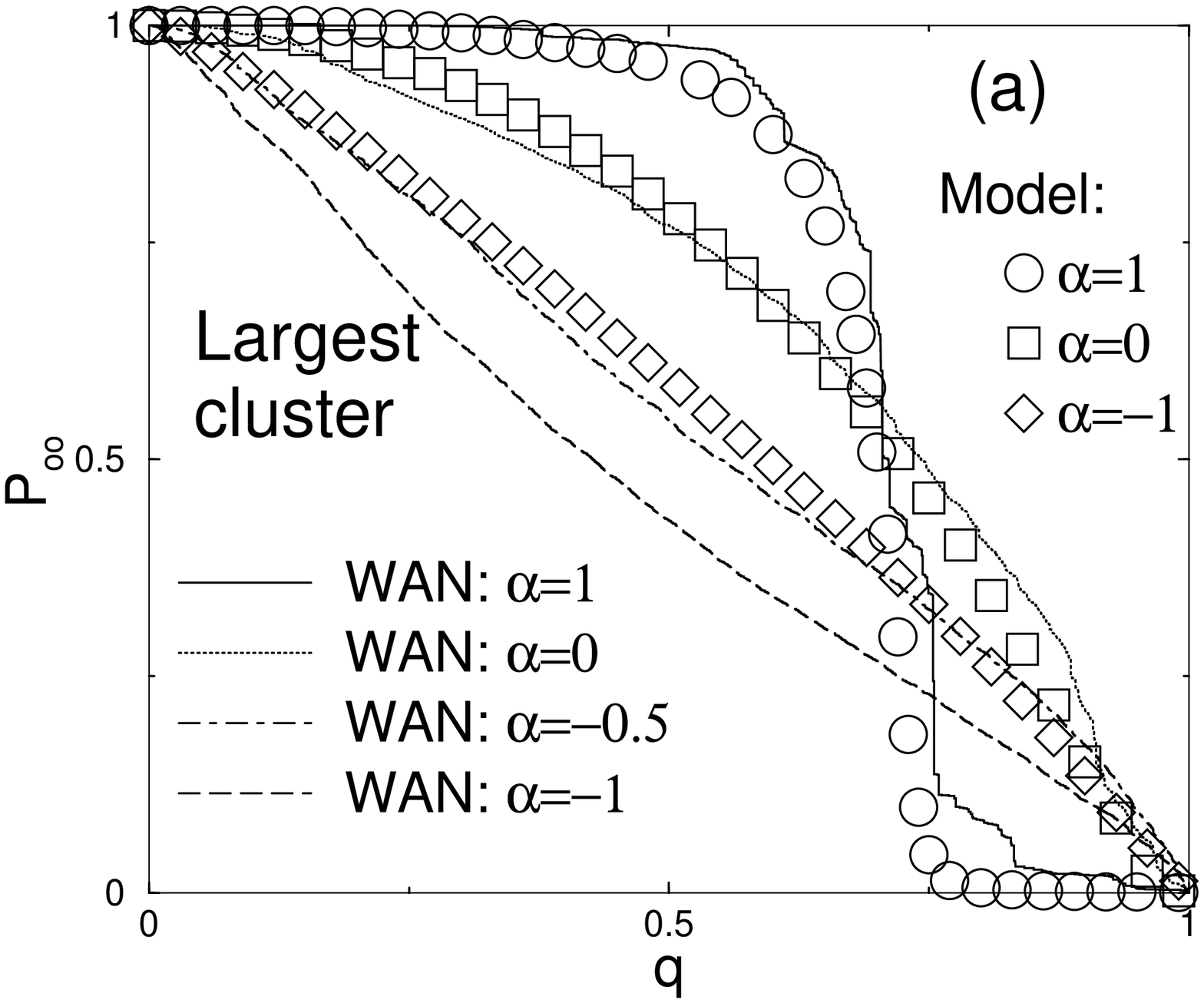}
  \includegraphics[width=0.37\textwidth, angle = -90]{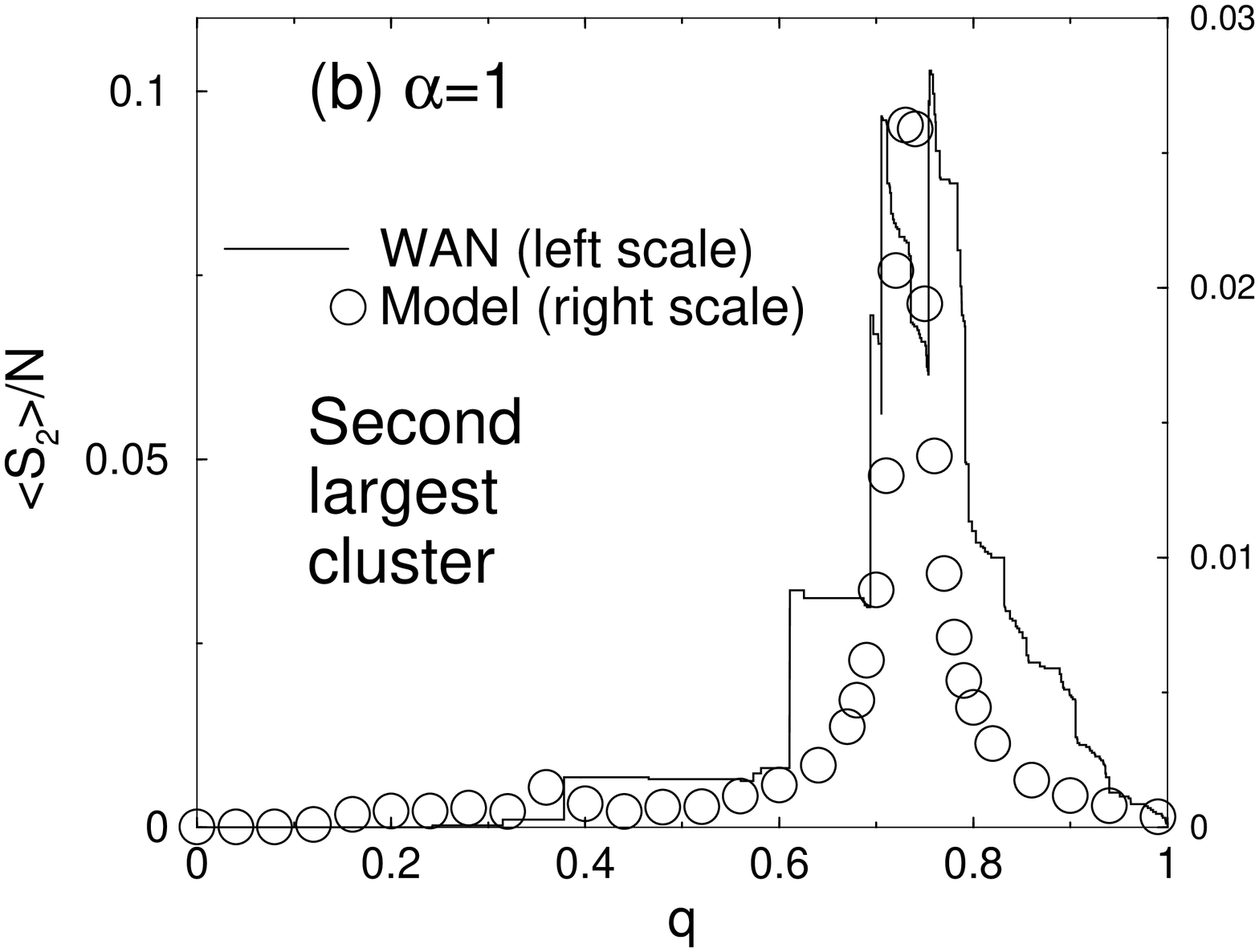}
  \includegraphics[width=0.37\textwidth, angle = -90]{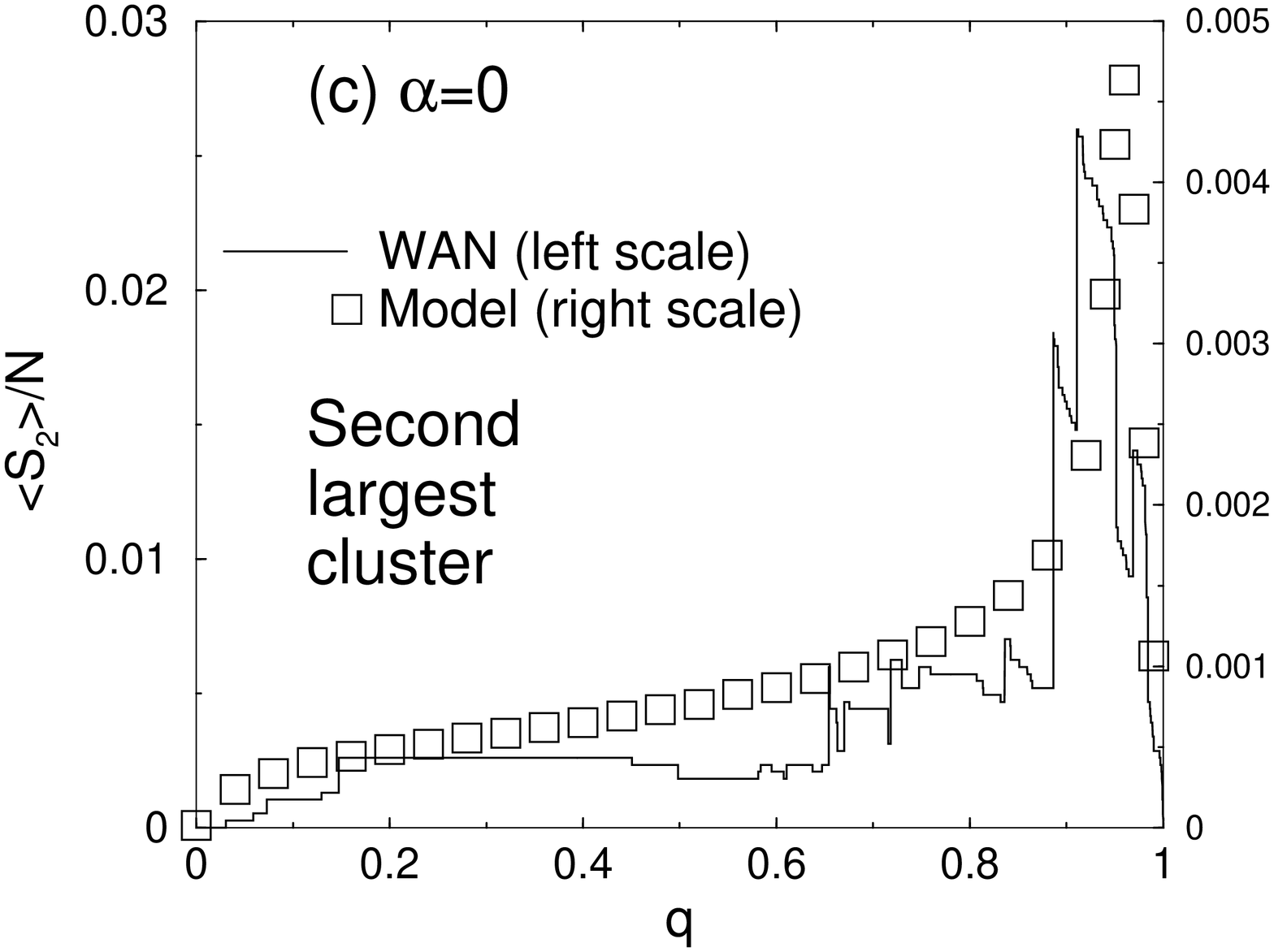}
  \includegraphics[width=0.37\textwidth, angle = -90]{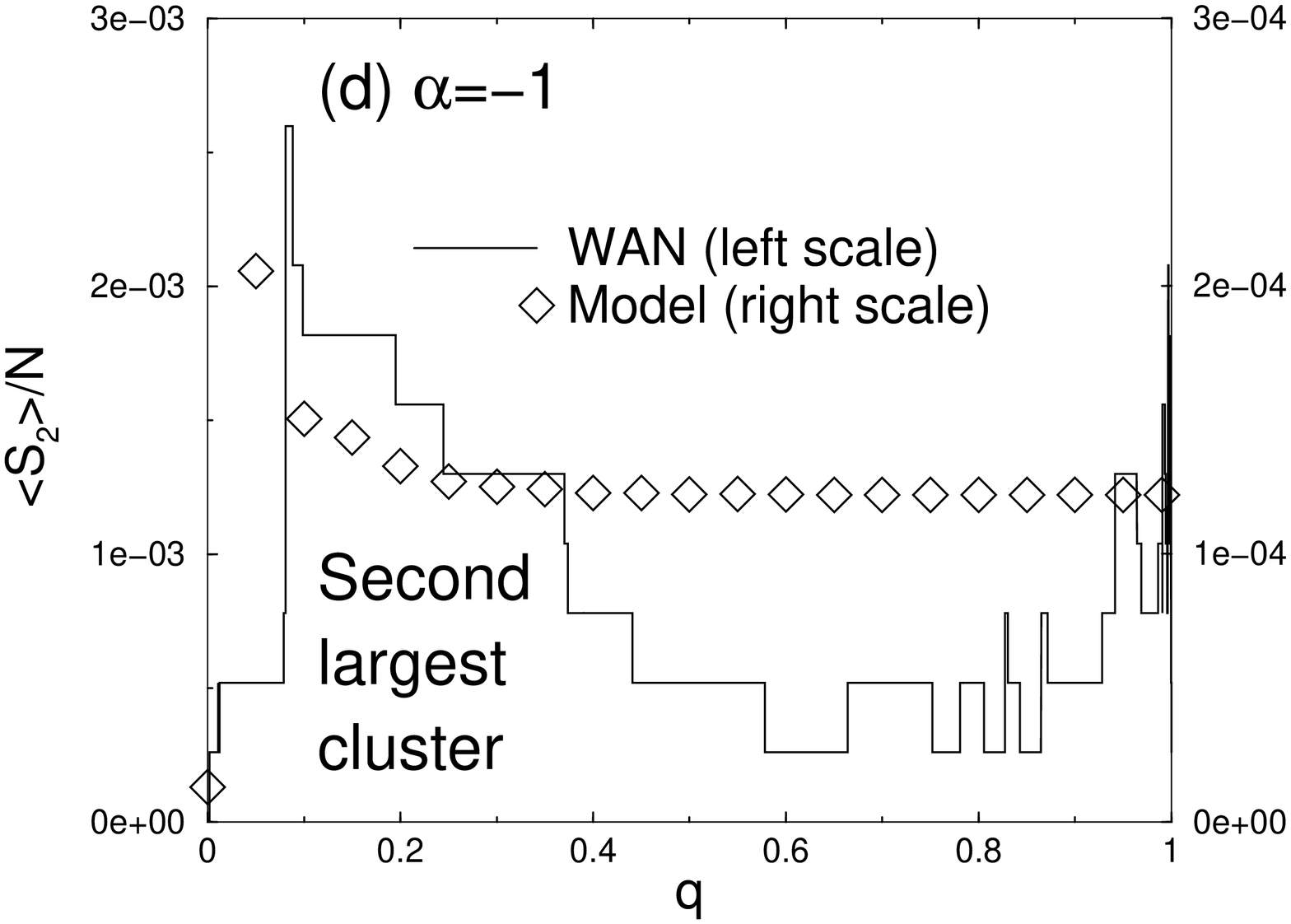}
  \caption{(a) $P_{\infty}$ as a function of $q$, the concentration of
    removed links in descending order of the weight for our model with
    $\lambda = 2.5$ (symbols) and the real WAN (lines). For model: $\alpha =
    1$ ($\bigcirc$), $\alpha = 0$ ($\Box$) and $\alpha = -1$
    ($\Diamond$). For the WAN network: $\alpha = 1$ (solid line), $\alpha =
    0$ (dotted line), $\alpha = -0.5$ (dot-dashed line) and $\alpha = -1$
    (dashed line). (b)-(d) $\langle S_{\rm 2} \rangle /N$ as a function of
    $q$ for our model with $\lambda = 2.5$ (symbols) and the real WAN (solid
    lines) with (b) $\alpha = 1$, (c) $\alpha = 0$ and (d) $\alpha = -1$.}
  \label{robust_cmp}
\end{figure}

Figure~\ref{robust_cmp} compares the robustness calculations (see also
Fig.~\ref{Robust} (a) and (d)) of our model with the real WAN. From the
calculation of both the largest and second largest clusters, we can see that
the real WAN networks have behavior similar to our model. Notice that
Fig.~\ref{robust_cmp} (a) shows $\alpha = -0.5$ is more robust than $\alpha =
-1$, which shows the advantage of $\alpha = -0.5$ in the real
WAN~\cite{FN_alpha}. The second largest cluster calculation (see also
Fig.~\ref{robust_cmp} (b)-(d) and Fig.~\ref{Robust} (d)) indicates where the
$p_c$ is, we see the WAN and our model are surprisingly similar.

\begin{figure}
  \includegraphics[width=0.6\textwidth, angle = -90]{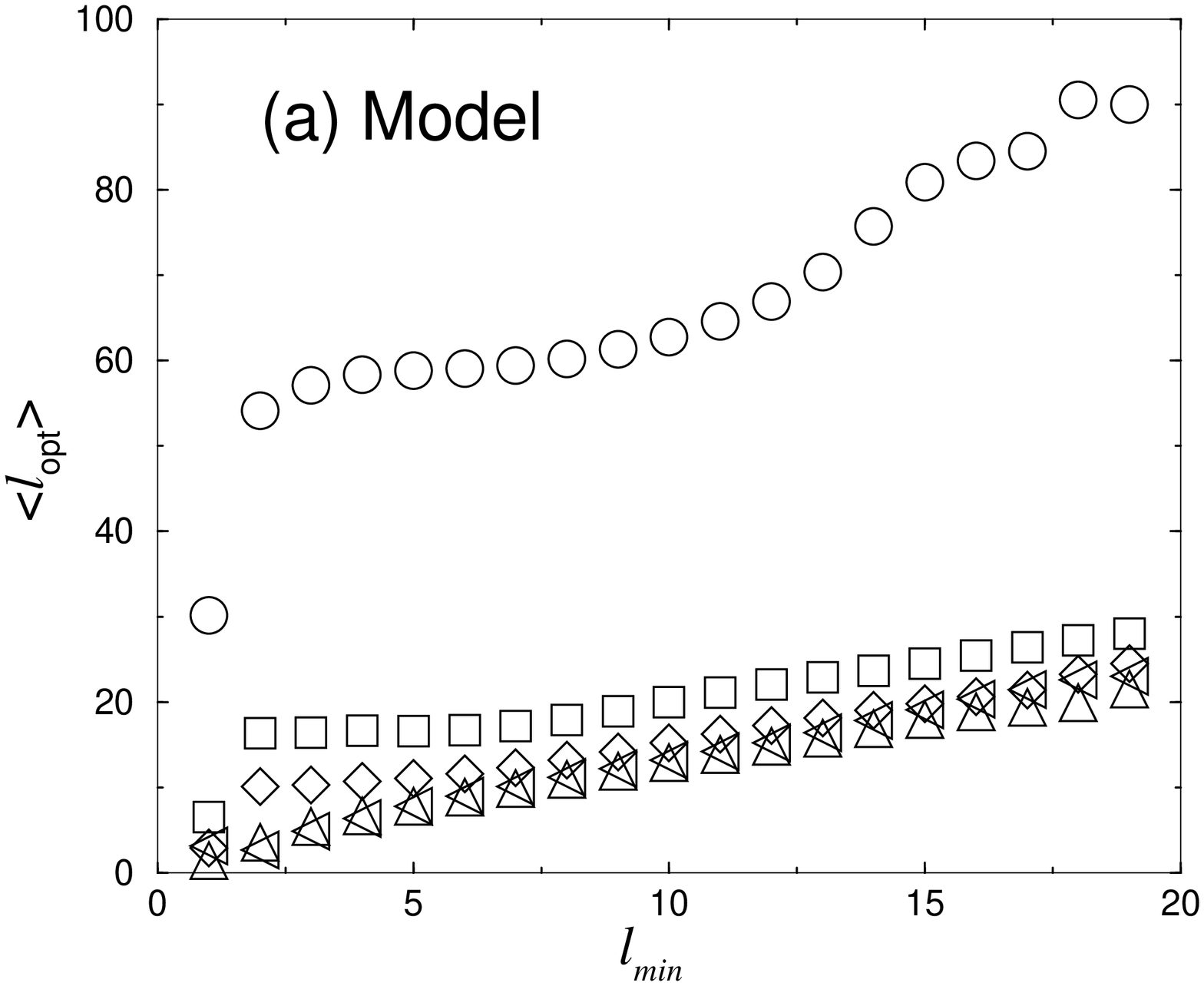}
  \includegraphics[width=0.6\textwidth, angle = -90]{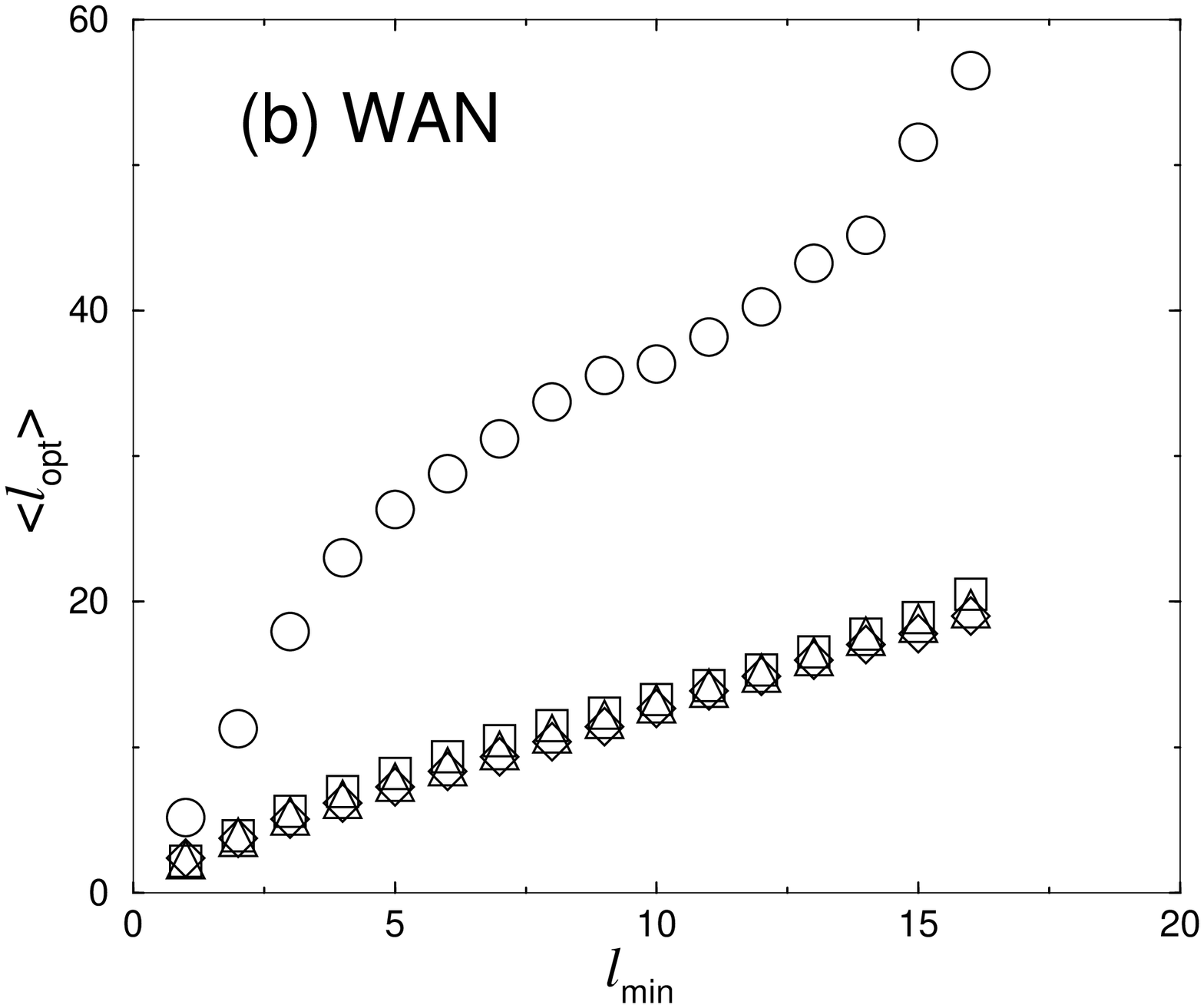}
  \caption{(a) Plot of $\langle \ell_{\rm opt} \rangle$ as a function of
    $\ell_{\rm min}$ for our model of correlated weighted SF network with
    $\lambda = 2.5$, in the SD limit ($\bigcirc$), $\alpha = 2$ ($\Box$),
    $\alpha = 1$ ($\Diamond$), $\alpha = 0$ ($\bigtriangleup$) and $\alpha =
    -1$ ($\lhd$). (b) Same plot as in (a) for the real WAN network with
    $\alpha = 1$ ($\bigcirc$), $\alpha = 0$ ($\Box$), $\alpha = -1$
    ($\Diamond$) and $\alpha = -0.5$ ($\bigtriangleup$).}
  \label{lopt_lmin_graph}
\end{figure}

To compare the optimal paths between the real WAN and our model, we calculate
$\langle \ell_{\rm opt} \rangle$ as a function of $\ell_{\rm
min}$. Figure~\ref{lopt_lmin_graph} shows the simulation results for the SF
model and for the real WAN (Fig.~\ref{lopt_lmin_graph} (b)). We observe that
for $\alpha \leq 0$, $\langle \ell_{\rm opt} \rangle$ is almost the same as
$\ell_{\rm min}$ because in the hub-philic regime the optimal paths tend to
go through the hubs, which shortens the length of the optimal paths. In the
hub-phobic regime ($\alpha > 0$), the optimal paths tends to avoid the hubs,
which naturally generate large length of the optimal paths. We also see that
in both the real WAN and the model for $\alpha > 0$, and for short $\ell_{\rm
min}$, $\langle \ell_{\rm opt} \rangle$ increases sharply, while for large
$\ell_{\rm min}$, $\langle \ell_{\rm opt} \rangle$ increases weakly with
$\ell_{\rm min}$. This is probably related to the crossover from strong
disorder in short length scales to weak disorder in large length scales that
was observed in several earlier studies~\cite{Porto, zhenhua, Sameet_mst}.

\section{Conclusions}

We propose a model of networks with correlated weights
(Eq.~(\ref{weight_def})) motivated by studies of the WAN and {\it E. Coli}
metabolic networks~\cite{Barrat_pnas, Macdonald_ecoli}. In our model, there
are three regimes according to the value of the parameter $\alpha$: the
hub-philic regime ($\alpha < 0$), the hub-phobic regime ($\alpha > 0$) and
the uncorrelated regime ($\alpha = 0$). We study the properties of the
optimal path and find two universality classes for the length of the optimal
path $\ell_{\rm opt}$: (i) For networks in the hub-phobic regime ($\alpha >
0$), the optimal path on SF networks with any $\lambda > 2$ behaves the same
as the optimal path in ER networks. (ii) For $\alpha \le 0$, the optimal
paths in SF networks belongs to the same universality class as the optimal
path in SF networks with uncorrelated weights.

We also calculate the robustness of the SF networks of our model and find that
SF networks in the hub-philic regime are more robust compared to SF networks
in the other two regimes. We propose an analytical method to study the
percolation transition of networks with weights of Eq.~(\ref{weight_def}) and
the numerical results suggest that the $p_c(\infty)$ of networks in the
hub-philic regime, although close to zero, is not zero for $3 < \lambda <
4$. We also observe strong finite size effects for SF networks
in the hub-philic regime.

In the last section, we compare the simulation results based on our model
with the actual WAN network. We calculate the $\langle \ell_{\rm opt} \rangle$
for two nodes separated with fixed distance $\ell_{\rm min}$ and observe
similar behaviors and the crossover from strong disorder to weak disorder for
both our model and the real WAN. We also compare the results of the
robustness calculation of our model with the real WAN network and find good
agreement.

\section{Acknowledgments}

The authors would like to thank Lazaros K. Gallos, Eduardo L\'{o}pez,
Gerald Paul and Sameet Sreenivasan for useful discussions and
suggestions. We thank Alessandro Vespignani for useful discussions and the
analysis of the anonymized WAN data that has been carried out in his
laboratory at the School of Informatics. We thank ONR, ONR-Global, UNMdP,
NEST Project No. DYSONET012911 and the Israel Science Foundation for support.

\newpage

\newpage

\newpage

\newpage

\newpage

\newpage

\newpage

\begin{thebibliography}{99}
\bibitem{FN_alpha} From the definition and the algorithm for the MST, only
  the relative order of the links according to the weight determines the
  structure of the MST. Thus from Eq.~(\ref{weight_def}), for all value of
  $\alpha > 0$ if $x_{ij}$ is fixed to be $1$, the same MST will be found for
  a specific structure of network (see also~\cite{Macdonald_ecoli}). In our
  simulation as we do many realizations with different series of random
  number $x_{ij}$, statistically results related the MST will be the same for
  all value of $\alpha$ with the same sign. For this reason, we only show the
  simulation results for $\alpha = 1, 0$ and $-1$. However, for a specific
  series of prefactor $x_{ij}$, different value of $\alpha$ with the same
  sign will give different MST. Thus while identifying the MST in the real
  WAN data, the value of $\alpha$ does matter.
\end{thebibliography}
\end{document}